\documentclass[a4paper]{article}
\usepackage{authblk}
\usepackage{booktabs}
\usepackage[dvipdfmx]{graphicx} 
\usepackage[top=30truemm,bottom=25truemm,left=25truemm,right=25truemm]{geometry}
\usepackage{dcolumn}
\usepackage{bm}

\usepackage{dsfont}
\usepackage{amsmath}
\usepackage{braket}
\usepackage{amsfonts}
\usepackage{amssymb}
\usepackage{color}
\usepackage{hyperref}
\hypersetup{colorlinks=true, citecolor= blue, linkcolor= blue, filecolor= blue, urlcolor= blue}
\usepackage{comment}
\usepackage{physics}
\usepackage{multirow}
\usepackage[dvipsnames]{xcolor}
\usepackage{lineno}
\usepackage[numbers, sort]{natbib}
\usepackage[normalem]{ulem}

\usepackage{orcidlink}
\newcommand{\bvec}[1]{{\bm{#1}}}
\usepackage{feynmp-auto}
\usepackage{adjustbox}
\usetikzlibrary{decorations.pathmorphing, arrows.meta}
\usepackage{caption}
\usepackage{subcaption}

\makeatletter
\patchcmd{\NAT@citex}
  {\@citea\NAT@hyper@\NAT@nmfmt{\NAT@nm}\@extra@b@citeb}
  {\@citea\NAT@nmfmt{\textit{\NAT@nm}}\@extra@b@citeb}
  {}{}
\makeatother

\title{Spin-lattice coupling induced chiral phonons and their signature in Raman Circular Dichroism}
\author[1,2,3]{Eduard Koller\orcidlink{0000-0002-4939-7470}  \footnote{eduard.koller@tum.de}}
\affil[1]{Technical University of Munich, TUM School of Natural Sciences, Physics Department, 85748 Garching, Germany}
\affil[2]{Institute for Advanced Study, Technical University of Munich, Lichtenbergstr. 2a, 85748 Garching, Germany}
\affil[3]{Munich Center for Quantum Science and Technology (MCQST), Schellingstr. 4, 80799 München, Germany}

\author[4]{Swetlana Swarup \footnote{swaru004@umn.edu}}
\affil[4]{School of Physics and Astronomy, University of Minnesota, Minneapolis, Minnesota 55455, USA}

\author[1,3,5]{Johannes Knolle}
\affil[5]{Blackett Laboratory, Imperial College London, London SW7 2AZ, United Kingdom}

\author[4,2]{Natalia B. Perkins}

\date{November 19, 2025} 

\begin{document}
\maketitle

\begin{abstract}

Recent Raman experiments on the Kitaev material $\alpha$-RuCl$_3$ have reported a finite Raman circular dichroism (RCD), revealing chiral phonon behavior not expected from lattice symmetry alone. To explain this observation, we develop a diagrammatic framework for the spin–phonon coupled Kitaev model. We demonstrate that bare phonons contribute no RCD, but coupling to the chiral spin excitation continuum under an applied magnetic field renormalizes the phonon propagator, mixing real polarization eigenvectors into complex superpositions with finite angular momentum. This interaction-induced modification generates a nonzero RCD accompanied by characteristic Fano line shapes in the Raman response, reflecting interference between discrete phonons and the continuum. The resulting signal grows with magnetic field strength, consistent with experiment, and directly tracks the field-induced chirality of the spin sector.
 More broadly, our results establish RCD as a powerful probe of interaction-induced chiral phonons in correlated quantum materials.

\end{abstract}

\section*{Introduction}

Phonons are among the most fundamental collective excitations in solids. Traditionally, they have been regarded as neutral bosons that transport heat and momentum but carry no intrinsic quantum numbers such as spin or angular momentum. In this conventional picture, vibrational eigenmodes are linearly polarized, and symmetry constraints such as time-reversal and inversion enforce the absence of angular momentum.
This perspective has evolved substantially over the past decade. Drawing inspiration from advances in topological insulators and semimetals, it is now recognized that phonons can carry angular momentum and even host nontrivial band topology. Phonon modes with finite angular momentum and 
nonzero helicity, known as chiral phonons, have been the subject of recent reviews \cite{Chen20182dchiralphononrev,Wang2024,Zhang2025SymmetryConstraints,Zhang2025rev,Juraschek2025}. 
Their emergence is closely tied to mechanisms such as spin–phonon coupling, which introduces Berry curvature into phonon bands and enables macroscopic transport phenomena, including the phonon Hall effect, where a magnetic field induces a transverse phonon heat current \cite{Strohm2005,Kagan2008,Zhang2010,Qin2012}.

Complementary to these transport signatures, intrinsic realizations of chiral phonons have also been observed in non-centrosymmetric materials such as monolayer transition-metal dichalcogenides, where broken inversion symmetry plays a crucial role \cite{Zhu2018ChiralPhonons,Chen20182dchiralphononrev,Wang2024}.
 In these systems, time-reversal symmetry remains intact, but the lack of inversion symmetry allows phonon modes belonging to two-dimensional irreducible representations to form circularly polarized combinations with finite  angular momentum   at high-symmetry points of the Brillouin zone. This symmetry-enabled mechanism underlies the valley-contrasting chiral phonons detected by ultrafast spectroscopy \cite{Zhu2018ChiralPhonons}. These observations provide direct evidence that crystalline symmetries can dictate whether phonons acquire chirality, a point of central importance when comparing non-centrosymmetric systems with centrosymmetric magnets discussed below.

Most theoretical and experimental studies of chiral phonons to date have worked within a single-particle framework \cite{Zhang2015HexagonalLattice,Ishito2022,Wang2024}. In this picture, chiral phonons are understood as circularly polarized vibrational modes that arise from doubly degenerate phonons at high-symmetry points of the Brillouin zone, 
transforming as two-dimensional irreducible representations of the crystal point group. The corresponding eigenvectors can be chosen as a pair of linearly polarized modes with orthogonal polarization vectors, $\boldsymbol{\epsilon}_{1}$ and $\boldsymbol{\epsilon}_{2}$. By forming complex linear combinations,  
$\boldsymbol{\epsilon}_{\pm} = \frac{1}{\sqrt{2}} \left( \boldsymbol{\epsilon}_{1} \pm i \boldsymbol{\epsilon}_{2} \right),
$
one obtains circularly polarized phonons that carry angular momentum $\pm \hbar$ along the mode axis. This single-particle picture has been widely applied to interpret the valley-selective chiral phonons observed in non-centrosymmetric crystals such as monolayer transition-metal dichalcogenides \cite{Zhu2018ChiralPhonons}.

An important experimental probe of chiral phonons is Raman circular dichroism (RCD), defined as the difference in Raman scattering intensity between left- and right-circular polarizations. Also referred to as helicity-resolved Raman scattering, RCD has been widely used to characterize phonon chirality at the Brillouin-zone center \cite{Chen2015,Ishito2022,Zhang2023,Yang2024,Kumar2024}. A robust experimental signature is the splitting between the $+-$ and $-+$ circularly polarized Raman channels, consistently observed and reproduced by theoretical modeling \cite{Zhang2025SymmetryConstraints,Zhang2023WeylPhonons,Ishito2022}. These results rely heavily on first-principles methods, including density functional theory (DFT) \cite{Ishito2022}, which capture the polarization dependence of Raman-active chiral phonons with good agreement to experiment.

While first-principle approaches successfully describe intrinsic realizations of chiral phonons in non-centrosymmetric crystals, they are limited to the single-particle picture. In centrosymmetric magnets, by contrast, phonons can acquire chirality only through many-body coupling to other chiral quasiparticles.
 A prominent example is provided by Kitaev spin liquids \cite{Kitaev2006,Knolle2017ARCMP,Takagi2019NRP,Motome2020,Trebst2022}, where spin excitations are fractionalized and can be described as continua of emergent Majorana fermions. The latter, and in turn the spin excitation continuum, acquire chirality once time-reversal symmetry (TRS) is broken. Crucially, we show that their coupling to phonons \cite{Ye_2020} hybridizes vibrational modes into complex superpositions, thus endowing them with finite angular momentum. 
 A complementary route arises from Raman-type spin–phonon interactions, where Berry-phase corrections to the Born–Oppenheimer approximation generate emergent gauge fields acting on phonons \cite{Saito2019}. In the Kitaev model, this interaction-driven mechanism is further enriched by field-induced scalar spin chirality, which skew-scatters chiral phonons \cite{Oh2025}. Together, these many-body effects not only modify phonon polarization vectors but also produce a finite RCD, unexpected from the symmetry of the underlying lattice. Moreover, the RCD response is predicted to scale with the strength of the applied magnetic field and to exhibit sensitivity to magnetic phase transitions, making it a powerful probe of correlated spin–phonon dynamics.

Recent Raman experiments on the Kitaev candidate $\alpha$-RuCl$_3$ provide striking evidence for this mechanism. 
When a magnetic field is applied along the crystallographic ${\bf a}$-axis, the crystal symmetry is reduced from $D_{3d}$ to $C_{2h}$ and TRS is broken. In this configuration,
 the bare phonons are constrained to carry zero angular momentum. Yet, a finite RCD has been observed by Anuja Sahasrabudhe {\it et al.}~\cite{sahasrabudhe2024,Sahasrabudhethesis2024}.
This finding is consistent with the idea that phonons acquire chirality through coupling to the chiral Majorana continuum once a magnetic field breaks TRS~\cite{Ye_2020,singh2024}. Motivated by these results and by related phenomena such as  observations of the thermal Hall effect in $\alpha$-RuCl$_3$ \cite{kasahara2018thermal,Bruin2021,Czajka2021}
and theoretical discussions of phonon Hall viscosity from phonon–spinon interactions \cite{VinklerAviv2018,Ye2021}, we develop a diagrammatic framework demonstrating how phonons can acquire chirality through many-body interactions. Within this framework, vibrational modes are renormalized into complex superpositions with finite angular momentum, leading to a finite RCD at finite magnetic fields in agreement with experiment. More generally, the same mechanism applies whenever phonons couple to chiral quasiparticles, yielding renormalized chiral phonons and helicity-dependent Raman responses beyond the expectations of lattice symmetry  of the bare system alone.

 In this work, we focus on the  RCD signal associated with the phonon peaks, while the broad background RCD originating from the magnetic continuum without phonons has been analyzed previously in Ref.~\cite{Koller2025}. Before turning to the detailed analysis, we begin with Fig.~\ref{fig:intro_figure}, which provides a compact summary of the key physical mechanisms explored in this work.
Panel (a) illustrates the experimental geometry: circularly polarized light probes the system via Raman scattering under a magnetic field applied along ${\bm a}$, which lowers the symmetry and enables new couplings between lattice vibrations and spin excitations.
 The figure also sketches the honeycomb lattice structure relevant for Kitaev materials like $\alpha$-RuCl$_3$, highlighting both the nearest-neighbor spin interactions and the additional field-induced terms (proportional to $\kappa$) that break symmetry and give rise to chirality in the spin sector.
Panel (b) compares the measured and calculated field-dependent RCD defined as the difference in scattered intensity between right- and left-circularly polarized light, $I^{\rm RCD}(\Omega)\equiv I^{+-}(\Omega)-I^{-+}(\Omega)$, where $\Omega=\omega_{\text{out}}-\omega_{\text{in}}$ is the Raman energy shift. A more quantitative discussion of the fitting procedure and parameter choices is presented in the Results section.
At zero magnetic field, $\kappa=0$, the response is symmetric: phonons are purely real, non-chiral vibrations, and the circular channels are indistinguishable, resulting in vanishing circular dichroism. However, once the field  $B$ is turned on, leading  to $\kappa\neq 0$, the spin liquid becomes chiral and the phonons no longer remain isolated. They hybridize with the spin excitations, forming complex, angular-momentum-carrying modes. This hybridization produces a characteristic, field-dependent RCD that reflects the underlying chiral structure of the system.
Taken together, the two panels illustrate the core mechanism we uncover: a magnetic field induces chirality in the spin sector, which is then transferred to the phonons through their mutual coupling. The result is a finite circular dichroism in Raman scattering, a striking and experimentally accessible signature of emergent chiral phonons in correlated quantum materials.

The  rest of the paper is organized as follows. We begin by introducing the spin–phonon coupled model that forms the basis of our analysis. We then show how the renormalization from the spin-phonon coupling results in vibrational modes of complex superpositions with net angular momentum. Next, we show that   bare phonons yield zero RCD, while the inclusion of the spin–phonon coupling produces a finite RCD, serving as a clear signature of phonon chirality. Then we review the diagrammatic framework for Raman scattering in the spin–phonon coupled systems and use it to calculate the RCD response. Finally, we present the results, highlighting the distinct contributions to the overall RCD signal and compare our theory to recent experimental data from Ref.~\cite{sahasrabudhe2024,Sahasrabudhethesis2024}.

\section*{Results}
\subsection*{Model}

\begin{figure}[h]
    \centering
    \includegraphics[width=\linewidth]{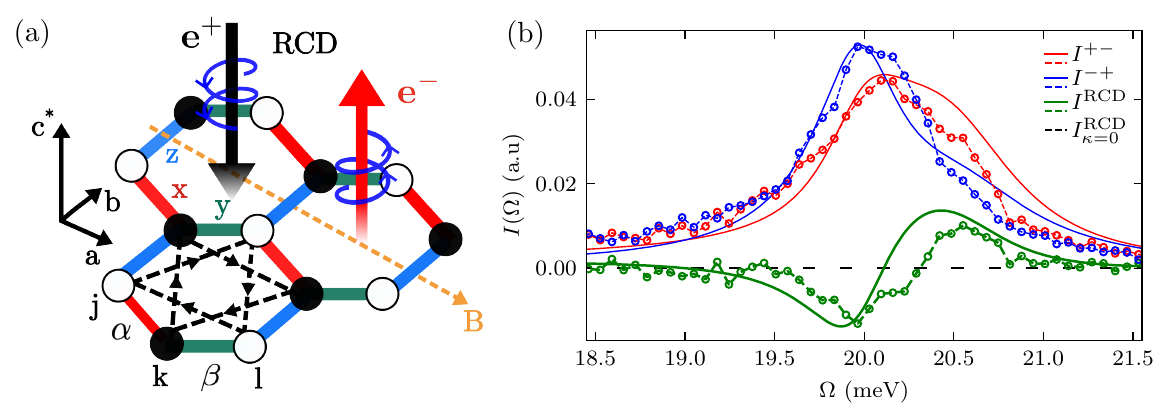}    \caption{\textbf{Experimental setup and Raman circular Dichroism} (a) $\bar{z} (\sigma^+\sigma^-)z$ geometry of the Raman scattering experiment. 
    In the monoclinic structure of $\alpha$-RuCl$_3$ the crystallographic $\bf c$-axis is tilted relative to the laboratory $\bf z$-axis \cite{Mai2019}, reducing the point group symmetry from $D_{3d}$ to $C_{2h}$. 
    Application of a magnetic field along the crystallographic $\bf a$-axis further lowers the symmetry to the magnetic point group $2^\prime/m^\prime$. Nearest-neighbor (NN) Majorana hopping amplitudes are indicated by the colored bonds, while the black star denotes the next-nearest-neighbor (NNN) Majorana hopping generated by the time-reversal symmetry breaking Zeeman term.  
    (b) Comparison between experimental RCD spectra at $B = 12.5$T (open circles, dashed lines as guides to the eye) and theoretical predictions at $\kappa$ = 0.18 (solid lines), highlighting the amplitude difference and frequency splitting between the $I^{+-}$ and $I^{-+}$ scattering channels. The comparison is carried out using the parameters listed in Tab.~\ref{tab:thyvsexp_parameters}. 
    At zero magnetic field, the absence of chiral excitations results in a vanishing RCD signal, indicated by a dashed black line.}
    \label{fig:intro_figure}
\end{figure}

We consider the minimal theoretical model that captures the essential coupling between the spinon continuum and  the lattice vibrations in $\alpha$-RuCl$_3$. The system is described by the Hamiltonian
\begin{align}\label{eq:model}
    H = H_\text{sp} + H_\text{ph} + H_\text{sp-ph},
\end{align}
which represents a spin-phonon coupled Kitaev model on a two-dimensional honeycomb lattice, subject to an external magnetic field.

\subsubsection*{Spin Hamiltonian  and TRS breaking}

The spin Hamiltonian $ H_{\text{sp}}$ is given by the isotropic Kitaev interaction with a TRS breaking perturbation
\begin{align} 
\label{eq:khm_h}
       H_{\text{sp}} = - J \sum_{\langle i j \rangle_\alpha  }  \sigma_i^\alpha \sigma_j^\alpha 
    - \kappa \sum_{ \substack{ \langle jkl \rangle}_{\alpha \beta}  }  \sigma_j^\alpha \sigma_k^\gamma \sigma_l^\beta,
\end{align}
where bonds are denoted as $\langle ij\rangle_\alpha$ and $\langle jkl \rangle_{\alpha \beta}$ represents path consisting of $\langle jk\rangle_\alpha$ and $\langle kl\rangle_\beta$ where $(\alpha \beta \gamma)$ is a permutation of $(x y z)$. The first term describes the bond-dependent Kitaev interactions, while the second term captures the leading perturbative TRS-breaking term from a  magnetic field via an effective three-spin interaction. Importantly, this three-spin term preserves the exact solvability of the model \cite{Kitaev2006}. 

The effective interaction strength $\kappa$ arises at third order in perturbation theory from the Zeeman coupling to a magnetic field $\mathbf{h} = (h_x, h_y, h_z)$, scaling as $\kappa \sim h_x h_y h_z / J^2$. 
Motivated by Raman experiments on $\alpha$-RuCl$_3$ 
\cite{sahasrabudhe2024,Sahasrabudhethesis2024}, we consider an in-plane magnetic field aligned with the crystallographic $\bf a$-axis, represented as $\mathbf{h} = (1,1,-2)$ in units of $J$. A finite $\kappa$ opens a topological gap in the spinon spectrum at the Dirac points, with magnitude $\Delta \sim 6\sqrt{3}\,\kappa$.

The model is solved by fractionalizing spins into Majorana fermions $\sigma^\alpha_i = i b_i^\alpha c_i$, which give rise to matter and flux excitations. The physical ground state resides in the flux-free sector \cite{Kitaev2006, Lieb1961}, reducing the problem to a quadratic Hamiltonian in matter fermions. It is diagonalized after a Fourier and  a Bogoliubov transformation to \begin{align}\label{spindiag}
    H_{\text{sp}} = 
\sum_\bvec{k} E(\bvec{k}) \left( a^\dagger_\bvec{k} a_\bvec{k} - 1/2  \right),
\end{align}
with $a^\dagger_{\bm{k}}$ the complex fermion creation operator 
and $E_{\bvec{k}}$ the  dispersion. $G(\tau,{\bm{k}}) = - \langle T_\tau a_{\bm{k}}(\tau) a^\dagger_{\bm{k}}(0) \rangle$ is the corresponding Matsubara Green's function of the complex matter fermions which is given in the Fourier space as
\begin{equation}\label{eqn:greenfunc}
    G^{(0)}(i\omega_n,\bvec{k}) = (i\omega_n - E_{\bvec k})^{-1}.
\end{equation}

\subsubsection*{Phonons and dynamical matrix}

The second term in Eq.~\eqref{eq:model} describes the free lattice dynamics.  Before focusing on the zone-center Raman phonons, we adapt the standard harmonic description. Using $\mu$ as the branch index of the phonon modes and $i =x,y,z$ for the cartesian components, the eigenvalue problem of the mass-weighted dynamical matrix $d^0(\bm q)$ is 
\begin{align}
    \sum_{\mu^\prime } d^{0}_{\mu \mu^\prime} (\bm q) \bm \epsilon_{\bm q,\mu^\prime} = \omega^2_{\mu}(\bm q) \bm \epsilon_{\bm q,\mu},  && \sum_i \epsilon_{i, \bm q, \mu}^* \epsilon_{i,\bm q,\mu^\prime} = \delta_{\mu \mu^\prime} .
\end{align}
This defines the polarization vectors $\bm \epsilon_{\bm q, \mu}$ and  frequencies $\omega_\mu(\bm q)$, yielding the spectral decomposition  
\begin{align}
d^0(\bm q)=\sum_{\mu}\omega_\mu^2(\bm q)\,|\bm \epsilon_{\bm q, \mu}\rangle\langle \bm \epsilon_{\bm q, \mu}|.
\end{align}
The free phonon Green’s function is the resolvent of the dynamical matrix $d^0(\bm q)$:
\begin{align}
D^{0}(\bm q,\omega)
= \big[\omega^2\mathds{1}-d^0(\bm q)\big]^{-1}
= \sum_{\mu}\frac{|\bm \epsilon_{\bm q, \mu}\rangle\langle \bm \epsilon_{\bm q, \mu}|}{\omega^2-\omega_\mu^2(\bm q)}.
\label{eq:free-propagator}
\end{align}
The Matsubara form follows by analytic continuation $\,\omega\!\to\! i\omega_m\,$. 
We will use both notations below.

Since Raman scattering primarily probes Brillouin-zone–center phonons, we now set $\bm q=0$ and, for the relevant optical modes, approximate them as 
\begin{align}
\omega_\mu(\bm q)\approx \omega_\mu(0)\equiv \omega_\mu^0.
\end{align}
At the zone center, the normal modes can be organized by point-group irreducible representations (irreps), which we label by $\Gamma$. 
We will suppress the $\bm q=0$ label when unambiguous. 
The diagonalized phonon Hamiltonian for the zone-center modes reads
\begin{align}
H_\text{ph}=\sum_{\Gamma}\omega_\Gamma^0\Big(b^\dagger_\Gamma b_\Gamma+\tfrac{1}{2}\Big),
\end{align}
and the (dimensionless) displacement in irrep $\Gamma$ is expanded as
\begin{align}
u_\Gamma(t)=\bm \epsilon_\Gamma\, b_\Gamma\, e^{-i\omega_\Gamma^0 t}
+\bm \epsilon_\Gamma^{*}\, b_\Gamma^\dagger\, e^{+i\omega_\Gamma^0 t},
\end{align}
with polarization vectors $\bm \epsilon_\Gamma$ normalized within the unit cell.
The corresponding Matsubara propagator is
\begin{align}
D^{0}_{\Gamma\Gamma'}(\tau,\bm 0)=-\big\langle T_\tau\,u_\Gamma(\tau)\,u_{\Gamma'}(0)\big\rangle,
\qquad
D^{0}_{\Gamma\Gamma'}(i\omega_m)=
\frac{2\,\omega_\Gamma^0}{(i\omega_m)^2-(\omega_\Gamma^0)^2}\,\delta_{\Gamma\Gamma'}.
\label{D0}
\end{align}

In an ideal two-dimensional honeycomb lattice the symmetry is $D_{3d}$, but experiments show that $\alpha$-RuCl$_3$ is monoclinic ($C2/m$), with interlayer tilts reducing the point group to $C_{2h}$ \cite{Johnson2015_RuCl_crystall_structure,Cao2016_RuCL_crystal_structure}. 
This fixes the allowed BZ-center phonon irreps and is central to the spin–phonon analysis. 
For the 24-dimensional vibrational space (two Ru and six Cl per unit cell, three components each),
\begin{align}
\Gamma_{\text{vib}} = 6 A_g \oplus 6 B_g \oplus 5 A_u \oplus 7 B_u,
\end{align}
where only the even irreps $A_g$ and $B_g$ are Raman active.
A magnetic field parallel to the crystallographic $\bf a$-axis (along $[{\bar 1}{\bar 1}2]$) further reduces the symmetry to the magnetic point group 
$2^\prime/m^\prime = C_{2h}(C_i)=\{\mathds{1},\, i,\, C_{2}\mathcal{T},\, \sigma_h \mathcal{T}\}$,
where $\mathds{1}$ is identity, $i$ inversion through a honeycomb center, $C_{2}$ the two-fold rotation around the crystallographic $\bf b$-axis, $\sigma_h$ reflection in the plane, and $\mathcal{T}$ time reversal.

\subsubsection*{Spin–phonon coupling}

 The third term in Eq.~\eqref{eq:model}, $H_\text{sp-ph}$, describes the spin–phonon coupling. Microscopically, this coupling originates from bond-length modulations induced by lattice vibrations, which in turn modify the Kitaev exchange $J$, thereby generating an effective spin–phonon interaction (schematically represented in Fig.~\ref{fig:diagram_panel}~(c))
\begin{align}
    \label{eq:spin-phonon-coupling}
    H_\text{sp-ph} = \sum_{\Gamma} \lambda_{\Gamma} \Sigma_{\Gamma} u_{\Gamma},
\end{align}
where $\lambda_{\Gamma} \sim \mathrm{d}J/\mathrm{d}\bvec{r}$ is the spin-phonon coupling constant (potentially distinct for different irreps), $u_{\Gamma}$ is the phonon displacement,
and $\Sigma_{\Gamma}$ the spin bilinear transforming as irrep $\Gamma$. For simplicity, we neglect phonon-induced generation of additional exchange interactions (e.g., Heisenberg-type terms), as these are expected to be  small. 
For the $C_{2h}$ point group, the spin bilinear transforming under the Raman-active irreps are \cite{singh2024}
\begin{align}\label{sigmaA}
\Sigma_{A_{g}} &=  \sum_\bvec{r} \left(\sigma_\bvec{r}^x \sigma_{\bvec{r}+\bvec{d}_x}^x + \sigma_\bvec{r}^y \sigma_{\bvec{r}+\bvec{d}_y}^y -2 \sigma_\bvec{r}^z \sigma_{\bvec{r}+\bvec{d}_z}^z\right),\\
\Sigma_{B_{g}} &=  \sqrt{3} \sum_\bvec{r} \left(\sigma_\bvec{r}^x \sigma_{\bvec{r}+\bvec{d}_x}^x - \sigma_\bvec{r}^y \sigma_{\bvec{r}+\bvec{d}_y}^y\right),
\label{sigmaB}
\end{align}
where $\bvec{d}_x=\begin{pmatrix} \sqrt{3}/2 & 1/2 \end{pmatrix}^T$ and $\bvec{d}_y=\begin{pmatrix} -\sqrt{3}/2 & 1/2\end{pmatrix}^T$ are the vectors connecting neighboring lattice sites along the $x$ and $y$ bonds, as shown in Fig.~\ref{fig:intro_figure}~(a).

\subsection*{Phonon chirality in spin-phonon coupled systems}

Phonons in harmonic, nonmagnetic, centrosymmetric crystals with TRS are not expected to carry angular momentum. For a phonon mode with momentum $\bm q$ and branch index $\mu$, the angular momentum can be written as \cite{McLellan1988,Zhang2014,Zhang2025rev}
\begin{align}
\label{eq:phonon_am}
l_{\bm{q},\mu }^{z} = \hbar\,   \bm \epsilon_{\bm q,\mu }^\dagger \bigoplus_{n=1}^{N}\begin{pmatrix} 
0 & - i \\ i & 0 \end{pmatrix} \bm \epsilon_{\bm q,\mu}
\end{align}
 where $\bm \epsilon_{\bm q,\mu}$ is the polarization vector of the phonon branch $\mu$.  The $2\times 2$ antisymmetric matrix acts in the $(x,y)$ displacement subspace, and  the  $\bigoplus_{n=1}^{N}$ extends this action to all $n =1,..,N$ atoms  in the unit cell \cite{Zhang2014}. 

In crystals with TRS, the phonon eigenvectors can always be chosen real. This immediately implies that
  $l^z_{-\bm q,\mu} = - l^z_{\bm q,\mu}$, 
 so that when summed over all wavevectors and branches,  the total phonon angular momentum
\begin{align} \label{angularm}
J^z =\sum_{\bm q,\mu } l^z_{\bm q, \mu} \left[ n_B(\omega_{\bm q, \mu}) + \frac{1}{2}\right]  
\end{align} 
vanishes. 

For Raman scattering, the most relevant phonons are those at the BZ center, ${\bm q}=0$, because photons transfer negligible momentum. For these zone-center modes, TRS enforces  $l^z_{ \bm 0,\mu}=0$, so they carry no angular momentum.

Additional crystallographic constraints can further enforce this result. In centrosymmetric point groups such as
s $C_{2h}$, non-degenerate phonons are symmetry-forbidden from carrying angular momentum, since their eigenvectors can always be chosen real \cite{Zhang2025SymmetryConstraints}. Only in the presence of symmetry-protected degeneracies could one, in principle, form complex linear combinations with finite angular momentum  but at $\bm q=0$, even degenerate modes are constrained by TRS  to yield $l^z_{ \bm 0,\mu}=0$.

Breaking these symmetry constraints, most importantly inversion,
 allows phonons to acquire finite angular momentum. In non-centrosymmetric crystals this can give rise to non-degenerate modes with circular (or, more generally, elliptical) polarization, as observed in monolayer transition-metal dichalcogenides such as MoS$_2$ and WS$_2$, where chiral phonons at the BZ center have been both theoretically predicted and experimentally observed \cite{Chen2015,Zhang2015HexagonalLattice,Chen20182dchiralphononrev}. In centrosymmetric magnets, however, such chirality can only emerge through coupling to other degrees of freedom, like spin–phonon interactions. 
When TRS is broken, the coupling to the chiral spinon continuum renormalizes the phonon polarization vectors, making them complex-valued and thereby endowing phonons with finite angular momentum.
This renormalization is captured by the one-loop self-energy within the Dyson equation framework, shown schematically in Eq.~\eqref{fig:Dyson}.
At the Brillouin-zone center, the Dyson equation takes the matrix form
\begin{align} \label{Dyson}
D_{\Gamma \Gamma^\prime }(i\omega_m ) = 
D^{0}_{\Gamma \Gamma^\prime }(i\omega_m)\delta_{\Gamma\Gamma^\prime} + \Big[
D^{0}(i\omega_m )
\Pi(i \omega_m) 
D(i\omega_m)
\Big]_{\Gamma \Gamma^\prime }.
\end{align}

For concreteness, we focus on the optical phonons belonging to the doubly degenerate $E_g$ representation of the $D_{3d}$ point group.
Applying  magnetic field along the ${\bf a}$-axis lowers the symmetry and lifts this degeneracy, splitting the $E_g$ doublet into two distinct, non-degenerate modes, which we label as $A_g$ and $B_g$\footnote{ 
Note that upon lowering the symmetry from
 from $D_{3d}$ to $C_{2h}$, and further to its magnetic counterpart $C_{2h}(C_i)$, the two-dimensional irrep  $E_g^{1,2}$ first splits into $A_g\oplus B_g$, and once TRS is broken, these evolve into two one-dimensional magnetic corepresentations
  $DA_g^{1,2}$. 
In the literature, however, the split phonon doublet is conventionally labeled $A_g$ and $B_g$. To maintain consistency with common usage and to avoid confusion, we will call the phonon modes $A_g$ and $B_g$ instead of $DA_g^{1,2}$ throughout the rest of this paper.
}.

Projecting Eq.~\eqref{Dyson} onto the $\{A_g,B_g\}$ subspace yields a $2\times2$ phonon propagator that couples the $A_g$ and $B_g$ modes:
\begin{align}\label{D-1}
    D^{-1} (i \omega_m) = 
    \begin{pmatrix}
       \left[  D^{0}_{A_g}(i \omega_m) \right]^{-1} - \Pi_{A_g A_g}  (i \omega_m)  
       && - \Pi_{A_g B_g}(i \omega_m) \\ 
      - \Pi_{B_g A_g}(i \omega_m) && 
      \left[  D^{0}_{B_g}(i \omega_m) \right]^{-1} - \Pi_{B_g B_g}  (i \omega_m) 
    \end{pmatrix}.
\end{align}
Here $D^{0}_{A_g}(i\omega_m)$ and $D^{0}_{B_g}(i\omega_m)$  are the bare propagators
of the two non-degenerate modes, obtained as special cases of the general expression for $D^{0}$  in Eq.~\eqref{D0}.
In the absence of a magnetic field, both originate from the same degenerate doublet $E_g$, so their bare frequencies are identical,
  $\omega^0_{A_g} = \omega^0_{B_g}=\omega_0$.

  The diagonal elements of Eq.~\eqref{D-1} describe the self-energy renormalization of each mode, while the off-diagonal terms
 $\Pi_{A_g B_g}$  encodes their coupling.
 These terms are odd under TRS and therefore vanish in the $\kappa=0$ limit. In this case, the propagator becomes diagonal, the two modes decouple, and the phonons carry no net angular momentum, consistent with the group-theoretical restriction that $A_g$ and $B_g$ are real representations.
  Consequently, phonons remain achiral even with spin–phonon coupling, unless TRS is broken.
When TRS is lifted by the field, the off-diagonal structure of the full phonon propagator $D^{-1}(i\omega_m)$ reflects the lowered magnetic symmetry and allows for finite angular momentum modes.

Diagonalizing the full inverse propagator in Eq.~(\ref{D-1}) or, equivalently, the full retarded dynamical matrix $d(\omega)$ defined by $D^{-1}(\omega)\propto \omega^2\mathds{1}-d(\omega)$, from which $d_{\Gamma \Gamma^\prime}(\omega) = \omega_{\Gamma}^2 |\bm \epsilon_{\Gamma}\rangle \langle \bm \epsilon_{\Gamma}|+ 2 \omega_{\Gamma}\Pi_{\Gamma \Gamma^\prime}(\omega)$, yields the renormalized polarization vectors. 
After analytic continuation $i\omega_m\!\to\!\omega+i0^+$, the eigenfrequencies follow from $\det D^{-1}(\omega)=0$. 
To leading order, they are
\begin{align}\label{splitmodes}
&\omega^\text{ren}_{+}  \approx  \omega_0 + \text{Re}\Pi_{A_g A_g}(\omega_0)   - \text{Im} \Pi_{A_g B_g}(\omega_0)\\
&\omega^\text{ren}_{-}  \approx  \omega_0 + \text{Re}\Pi_{B_g B_g}(\omega_0)   +  \text{Im} \Pi_{A_g B_g}(\omega_0).
\end{align}
The resulting mode splitting is 
\begin{align}\label{split}
|\omega^\text{ren}_{+} - \omega^\text{ren}_{-}| \approx  2 \text{Im}\Pi_{A_g B_g}(\omega_0).
\end{align}
While these frequency shifts quantify the effect of the spin–phonon coupling, the polarization vectors of the renormalized modes are equally important, because they determine the angular momentum carried by the phonons. To obtain them, we must diagonalize the full retarded dynamical matrix.

\begin{figure*}
\center
\includegraphics[width=1\textwidth]{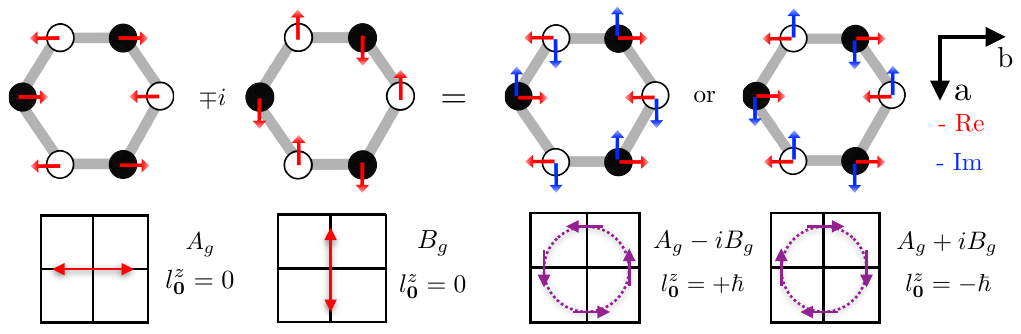}%
\caption{\label{fig:linear_circular_phonons} \textbf{Linear and circular phonon modes.}
TRS enforces real $A_g$ and $B_g$ phonon modes at the $\Gamma$-point, while the hybridization with the chiral-spinon continuum renormalizes the phonon modes into complex combinations with opposite angular momentum. 
}
\end{figure*}

We now consider the general situation in which the diagonal self-energies are equal,
  $\Pi_{A_gA_g}=\Pi_{B_gB_g}$ and the off-diagonal terms differ only by a phase, 
\begin{align}
    \Pi_{A_g B_g} = e^{i \phi_1} |\Pi_{A_g B_g}|, \quad \Pi_{B_g A_g} = e^{i \phi_2} |\Pi_{A_g B_g}|.
\end{align}
This form allows for complex-valued couplings between the  $A_g$ and $B_g$ modes, which is essential for generating phonon chirality.
In our model, the phases satisfy
 $\phi_1 = \phi_2 + \pi$.
Introducing the relative angle $\phi = \frac{1}{2}(\phi_1 - \phi_2)$, 
the phonon eigenvectors take the form
\begin{align}
\label{epsilonplusminus}
    \bm \epsilon_{\pm } = \frac{1}{\sqrt{2}}
    \left( 
        \bm \epsilon_{A_g}   \pm e^{-i\phi} \bm \epsilon_{B_g}
    \right).
\end{align}
To evaluate the corresponding angular momentum, we consider the polarization vectors in the two-Ru-atom subspace within the crystallographic $\bf{a}$-$\bf{b}$ plane, i.e $\bm \epsilon_{\Gamma} = \begin{pmatrix}a_1 & b_1 & a_2 & b_2 \end{pmatrix}^T $. From symmetry analysis \cite{Dresselhaus2008} the linear in-plane polarizations are 
\begin{align}
\label{eq:ru_polarization}
    \bm \epsilon_{A_g} = \frac{1}{\sqrt{2}} \begin{pmatrix} 0 & -1 & 0 & 1 \end{pmatrix} &&
    \bm \epsilon_{B_g} = \frac{1}{\sqrt{2}} \begin{pmatrix} -1 & 0 & 1 & 0 \end{pmatrix}, 
\end{align}
as illustrated in Fig.~\ref{fig:linear_circular_phonons}. The other Ru $B_g$ mode is out of plane, while the other Raman active irreps are Cl-lattice vibrations. 
Substituting the polarization vectors into Eq.~\eqref{eq:phonon_am}, the corresponding angular momentum is 
\begin{align}
    l_{\bm 0,\pm}^z = \pm \hbar \sin \phi .
\end{align}

For $\phi=0,\pi$, the resulting phonons are linear polarized ($l^z=0$), 
while $0<|\phi|<\pi/2$ gives elliptical polarization, and $\phi=\pm\pi/2$ is the circular polarized case ($|l^z|=\hbar$), as shown in Fig.~\ref{fig:linear_circular_phonons}. 
In this parameterization, the \emph{imaginary part} of  $\Pi_{A_gB_g}$ controls the mixing strength and the mode splitting, whereas the \emph{phase} $\phi$ fixes the handedness and magnitude of $l^z$. Thus, TRS breaking, which generates 
$ \Pi_{A_g B_g}\!\neq\!0 $, converts the two linear, \emph{achiral} zone-center modes into a pair with opposite helicities.

The total angular momentum at $\bm q=0$ follows from the renormalized polarizations. Using Eq.~(\ref{angularm}), we get
\begin{align}
J^z_{\bm 0}=\sum_{\Gamma =  \pm  } l^z_{\bm 0,\Gamma}\Big(\langle b^\dagger_{\bm 0,\Gamma}b_{\bm 0,\Gamma}\rangle+\tfrac{1}{2}\Big).
\end{align}

At $T=0$, the two chiral phonon modes carry opposite angular momenta ($+\hbar$ and $-\hbar$), so their zero-point contributions cancel and the ground state has $J^z_{\bm 0}=0$. 
At finite $T$, however,
a field-induced splitting of the doublet, Eq.~\eqref{splitmodes}, implies unequal Bose occupations $n_B(\omega_{+}^\text{ren})\neq n_B(\omega_{-}^\text{ren})$ and, in general,  unequal quasiparticle weights $Z_{+} \neq Z_{-}$. The latter are the residues of the renormalized phonon propagator in the  $\pm$ basis, 
$ Z_{\pm} = \left[ \partial D_{\pm}^{-1}(\omega)/\partial \omega \right]_{\omega = \omega_{\pm}^{\text{ren}}}^{-1}$, where
\begin{align}
D_{\pm}^{-1}(\omega)
= \frac{ \omega^{2} - \omega_{0}^{2}
- 2\omega_{0}\!\left[
    \mathrm{Re}\,\Pi_{A_gA_g}(\omega)
    \mp  \mathrm{Im}\,\Pi_{A_gB_g}(\omega)
\right]
- 2i\omega_{0}\!\left[
    \mathrm{Im}\,\Pi_{A_gA_g}(\omega)
    \pm  \mathrm{Re}\,\Pi_{A_gB_g}(\omega)
\right] }{2 \omega_0} 
\end{align}
 Linearizing $D_{\pm}^{-1}(\omega)$ around the poles and assuming  $\mathrm{Im}\,\Pi_{A_gA_g}(\omega)=0
    $ and $ \mathrm{Re}\,\Pi_{A_gB_g}(\omega)=0$,
      as supported by  Fig.~\ref{fig:polarization_bubble}, yields the complex residues
\begin{align}
Z_{\pm}
=\frac{\omega_0}{
\omega-\omega_0\partial_\omega \mathrm{Re}\Pi_{A_gA_g}(\omega)
\pm \omega_0\partial_\omega \mathrm{Im}\Pi_{A_gB_g}(\omega)
}\Bigg|_{\omega=\omega_{\pm}^{\text{ren}}},
\end{align}
which quantify the renormalization of the phonon amplitudes due to spin–phonon coupling.

Consequently, the thermal imbalance between the two circular branches produces a finite net angular momentum,
\begin{align}
J^z_{\bm 0}
=\hbar\sin\phi\,[Z_{+}\,n_B(\omega_{+}^\text{ren})-Z_{-}n_B(\omega_{-}^\text{ren})\,].
\end{align} 
If the occupations are nearly equal, 
$n_B(\omega_{+}^{\text{ren}}) \approx n_B(\omega_{-}^{\text{ren}})$,
one finds analytically that 
$J^z_{\mathbf 0} \propto \mathrm{Im}\,\Pi_{A_g B_g}$,
confirming that the finite angular momentum originates from the field-induced 
mixing of the orthogonal phonon components. 
A nonequilibrium imbalance, for example, induced by circularly polarized 
Raman excitation that selectively populates one branch, would likewise yield $J^z_{\bm 0}\neq 0$, even at zero temperature in the degenerate limit.

\subsection*{Raman Response}

 A detailed derivation of the Raman response  and its decomposition into spinon, phonon, and interference contributions is presented in the Methods section. Here we summarize the essential expressions and focus on their physical interpretation.

\begin{figure*}
\includegraphics[width=1\textwidth]{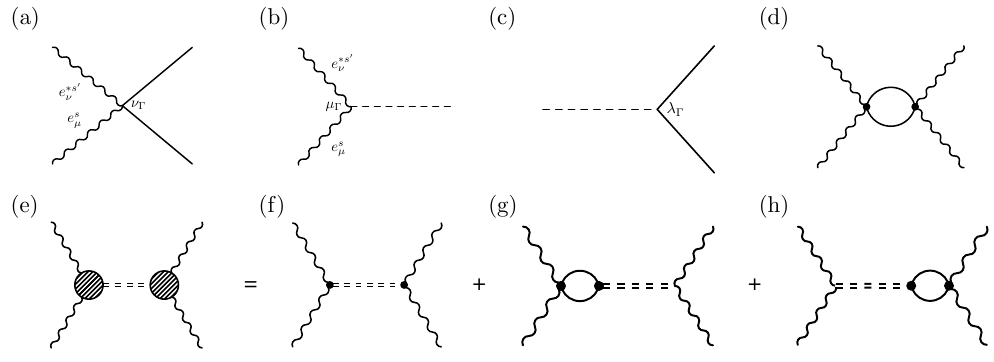}%
\caption{\label{fig:diagram_panel} \textbf{Diagramatic Raman scattering}
(a) Photon-spin vertex with spin-photon coupling constant $\nu$. (b) Photon-phonon vertex with phonon-photon coupling constant $\mu_{\Gamma}$. (c) Spin-phonon coupling vertex with spin-phonon coupling constant $\lambda_\Gamma$. (d) Magnetic Raman scattering. (e) Phonon scattering with renormalized phonon and renormalized vertex and its contribution in (f)-(h). (f) Phonon Raman scattering with renormalized phonon (g) left and (h) right Fano diagram. 
Note, that each vertex per diagram comes with a sum over the different irreps.}
\end{figure*}

The Raman intensity for a spin-phonon coupled system in the interaction picture is defined as
\begin{align} \label{eqn:intensityram}
I^{ss^\prime}(\Omega,\beta ) = \int d\mathrm{t} e^{i \Omega t} \left \langle \mathcal{T}_t 
         \left[R^{ss^\prime}(t)\right]^\dagger R^{s s^\prime}(0) e^{-i \int \mathrm{d}t^\prime H_\text{sp-ph}(t^\prime)}\right \rangle_\beta, 
\end{align} 
where $s,s^\prime$  are  incident/scattered polarizations, $\Omega=\omega_{\text{out}}-\omega_{\text{in}}$ is the Raman shift (energy transfer),
$\beta=1/T$
 and 
 $\langle ...\rangle_\beta = \text{Tr}[e^{-\beta H_0}]/\text{Tr}[e^{-\beta H_0}]$  is the thermal average over the Hilbert space of the spin-phonon Hamiltonian $H_0 = H_{\text{sp}} + H_\text{ph}$. 
Treating $H_{\text{sp-ph}}$ perturbatively, we expand the $S$-matrix
$\mathcal{S}=\mathcal{T}_t\exp\!\big[-i\!\int dt'\,H_{\text{sp-ph}}(t')\big]$ into the Dyson series, following Ref.~\cite{Feng_2022}. 
The diagrams involve two types of interaction vertices: 
(i) the spin–phonon coupling $H_{\text{sp-ph}}$ [Fig.~\ref{fig:diagram_panel}(c)], and 
(ii) the Raman vertex $R^{ss'}=R^{ss'}_{\text{em-ph}}+R^{ss'}_{\text{em-sp}}$, comprising the photon–phonon [Fig.~\ref{fig:diagram_panel}(b)] and photon–spin [Fig.~\ref{fig:diagram_panel}(a)] couplings.
The total Raman intensity is therefore
 \begin{align}\label{Fullintensity}
    I
    ^{ss^\prime} (\Omega,\beta) =  
    I_{\substack{\text{s}}}^{ss^\prime} (\Omega,\beta) + 
    I_{\substack{\text{ph}}}^{ss^\prime} (\Omega,\beta) +
    I_{\text{Fano,L}}^{ss^\prime} (\Omega,\beta) +
    I_{\text{Fano,R}}^{ss^\prime} (\Omega,\beta).
\end{align}
The spinon term $I_{\text{sp}}$ represents scattering from the fractionalized continuum, while $I_{\text{ph}}$ describes phonon peaks renormalized by the spin–phonon self-energy $\Pi(\omega)$ through
$D(\omega) = [D^{0}(\omega)^{-1} - \Pi(\omega)]^{-1}$.
The Fano terms encode photon–phonon and photon–spin mixing mediated by the continuum and yield the characteristic asymmetric lineshape.

We will work at zero temperature  ($\beta \to \infty$), where the flux sector remains in its ground state; thermally activated fluxes at finite $T$ introduce additional scattering processes that are beyond the scope of the present treatment.
Within this low-temperature framework, the different Raman vertices can be specified explicitly.

The phonon Raman operator describes the coupling of light to lattice vibrations through the  photon–phonon vertex:
\begin{align}
    \label{eq:def_phonon_R}
    R^{ss^\prime}_\text{em-ph} = \sum_{\Gamma} \mu_\Gamma \sum_{\mu ,\nu} e^{s}_\mu  e^{*s^\prime}_\nu r_{\text{em-ph},\Gamma}^{\mu \nu} u_{\Gamma},
\end{align}
where ${\bf e}^{s}$ and ${\bf e}^{s^\prime}$ are the polarizations of the in- and out-going photon, $\Gamma$  labels the irreducible representations of the (magnetic) point group, $\mu_\Gamma$ is the phonon-photon coupling strength, and $u_\Gamma$ are the phonon displacements transforming in the representation $\Gamma$.

In the following, we focus on the $E_g$ 
 phonon modes with energy $\approx 20.1 \,\text{meV}$,
which exhibit helicity-dependent Raman response  in $\alpha$-RuCl$_3$ ~\cite{sahasrabudhe2024,Sahasrabudhethesis2024}.
Monoclinic stacking lowers the symmetry, splitting the doubly degenerate $E_g$ representation into $A_g \oplus B_g$ modes.
The corresponding $C_{2h}$ Raman tensors are
\begin{align} 
r_{\text{em-ph},A_\mathrm{g}} = 
\begin{pmatrix}
           C &  0 & D \\
           0 & -C & 0 \\
           D &  0 & 0 
        \end{pmatrix}
 && 
r_{\text{em-ph}, B_\mathrm{g}} = 
\begin{pmatrix}
           0 & E & 0 \\
           E & 0 & F \\
           0 & F & 0 
        \end{pmatrix}, 
\end{align}
 with $C,D,F,E \in \mathbb{R}$. In this work, we restrict our analysis to the $({\bf a}$,${\bf b})$ crystallographic plane, corresponding to the upper 2×2 part of the Raman tensor. 

The spin Raman operator describes the coupling of light to magnetic excitations through the Loudon–Fleury mechanism~\cite{Fleury1968,Shastry1990,Shastry1991}
\begin{align}
    \label{eq:def_LF_R}
    R^{ss^\prime}_\text{em-sp} = \nu \sum_{\mu ,\nu} e^{s}_\mu e^{*s^\prime}_\nu \sum_{\bvec{r},\alpha } J_\alpha d_\alpha^\mu d_\alpha^\nu \sigma_{\bvec{r}}^\alpha \sigma_{r+\bvec{d}_\alpha}^\alpha,
\end{align}
where $\nu$ is the spin–photon coupling, $J_\alpha$ the Kitaev exchange, and $\bvec{d}_\alpha$ the bond vector. Analogous to the phonon case, the Raman tensor can be decomposed into irreducible representations,
\begin{align}R^{ss^\prime}_\text{em-sp} = \sum_\Gamma \alpha_\Gamma  \sum_{\mu ,\nu} e^{s}_\mu e^{*s^\prime}_\nu  r_{\text{em-sp},\Gamma}^{\mu \nu} \Sigma_\Gamma,
\end{align}
such that the coefficients $\alpha_\Gamma$ are determined from the orthogonality relations.

In the diagrammatic expansion of the spin–phonon coupled Raman response, the photon–phonon vertex becomes renormalized by the spinon continuum through the polarization bubble. This is illustrated 
in Fig.~\ref{fig:diagram_panel}  (e)-(h). 
As discussed in Ref.~\cite{Feng_2022}, this renormalization can be expressed in terms of left (L) and right (R) Raman vertices, which resum the corresponding diagrammatic series and naturally incorporate the interference between the spinon and phonon channels. This interference produces the Fano-type asymmetry in the Raman response and, through the off-diagonal self-energy $\Pi_{A_gB_g}$ and its phase, provides the microscopic mechanism for the finite Raman circular dichroism discussed in the next section.

\subsection*{Raman Circular Dichroism}
The RCD quantifies the difference in Raman intensities between left- and right-circularly polarized light. It is defined as
\begin{align}
\label{eq:def_RCD}
I^{\text{RCD}}(\Omega) =  I^{+-}(\Omega) - I^{-+}(\Omega),
\end{align}
where $\bvec{e}^+ = (1, i)^T/\sqrt{2}$ and $\bvec{e}^- = (\bvec{e}^+)^*$ denote the photon polarization vectors for left- and right-circular polarization. A finite RCD is a hallmark of TRS breaking, as the two polarization-resolved Raman intensities $I^{+-}(\Omega)$ and $I^{-+}(\Omega)$ are related by TRS. Following Eq.~(\ref{Fullintensity}), we analyze the RCD by separating the individual contributions from the spinon, phonon, and left- and right-Fano channels.

\begin{figure*}[t]
\includegraphics[width=1\textwidth]{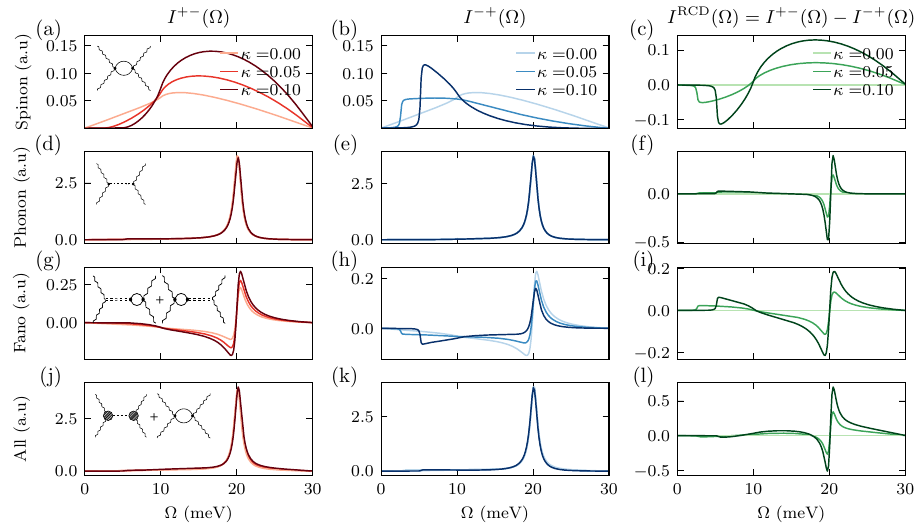}  %
\caption{\label{fig:rcd_overview} 
Raman response of a spin–phonon coupled Kitaev spin liquid. Rows show the separate 
contributions from spinons (a–c), phonons (d–f), Fano interference (g–i), and the total response 
(j–l). Columns correspond to the intensities $I^{+-}(\Omega)$ in the $\overline{z}(\sigma^+\sigma^-)z$ 
geometry, $I^{-+}(\Omega)$ in the $\overline{z}(\sigma^-\sigma^+)z$ geometry, and their difference 
$I^{\mathrm{RCD}}(\Omega) = I^{+-}(\Omega) - I^{-+}(\Omega)$. Insets in the upper-left corners show 
the corresponding Feynman diagrams. Numerical evaluation uses a $600\times600$ $\mathbf{k}$-mesh and 200 energy intervals over $\Omega\in[ 0,12J]$ with $J=10~\text{meV}$, the phonon broadening parameter in $i\Omega\to\Omega+i\delta$ is $\delta=0.04$.
}
\end{figure*}

The spinon contribution to the RCD, $I^{\text{RCD}}_{\substack{\text{sp} }} (\Omega)=I^{+-}_{\text{sp} } (\Omega)-I^{-+}_{\text{sp} } (\Omega)$,  
 calculated from the explicit expression in 
 Eq.~(\ref{Ispinfinal}) in  the {\it Methods},
is shown in Fig.~\ref{fig:rcd_overview}~(c) and Ref.~\cite{Koller2025}. 
As discussed therein, the RCD is intimately connected to the quantum geometry of the spinon bands: it is proportional to a  rescaled  single-particle density of states, vanishes inside the topological gap, and appears only once time-reversal symmetry is broken by a finite magnetic field.

The phonon contribution to the RCD, $I^{\text{RCD}}_{\substack{\text{ph} }} (\Omega)=I^{+-}_{\substack{\text{ph} }} (\Omega)-I^{-+}_{\substack{\text{ph} }} (\Omega)$,  with explicit expressions defined in Eq.~(\ref{Iphononfinal}) in  {\it Methods}, can be written as
\begin{align}
\label{eq:phonon_RCD}
 I^{\text{RCD}}_{\substack{\text{ph} }} (\Omega) &=  4  \Im \sum_{\Gamma,\Gamma^\prime} \mu_{\Gamma} \mu_{\Gamma^\prime } D_{\Gamma \Gamma^\prime} (\Omega) \nonumber \\ 
 &\times i \left[(r^{aa}_{\text{em-ph},\Gamma} -r^{bb}_{\text{em-ph},\Gamma}) \left( r^{ab}_{\text{em-ph},\Gamma^\prime} \right)^* - 
 r^{ab}_{\text{em-ph},\Gamma} \left( \left(r^{aa}_{\text{em-ph},\Gamma^\prime}\right)^* - \left( r^{bb}_{\text{em-ph},\Gamma^\prime} \right) ^*\right) 
 \right] ,
\end{align}
 where  the irrep-specific  constants $r^{aa}_{\text{em-ph},\Gamma}$, $r^{ab}_{\text{em-ph},\Gamma}$, and $r^{bb}_{\text{em-ph},\Gamma}$ are the matrix elements of the phonon Raman tensor $r_{\text{em-ph},\Gamma}$ in the crystallographic ${\bf a}$-${\bf b}$ basis.
When TRS is preserved, all elements of the Raman tensor are real, and the phonon RCD therefore vanishes.

Breaking TRS allows the tensor elements to acquire complex phases, which immediately generates a finite RCD signal. Such complex matrix elements can arise dynamically from vertex renormalization due to spin–phonon coupling, as illustrated in Fig.~\ref{fig:diagram_panel}, where the coupling to the spinon continuum mixes orthogonal phonon polarizations.
Alternatively, even in the absence of spin interactions, degenerate $E$-type phonons can form complex linear combinations
$\bm{\epsilon}^{\pm}$, as defined in Eq.~(\ref{epsilonplusminus}), corresponding to left- and right-circular lattice motions.
The allowed phonon angular momenta and the associated complex Raman tensor elements are fully determined by the magnetic point-group symmetry~\cite{Zhang2025SymmetryConstraints,Ovander1960,Cracknell_1969}.

\begin{figure*}[h]
\includegraphics[width=1\textwidth]{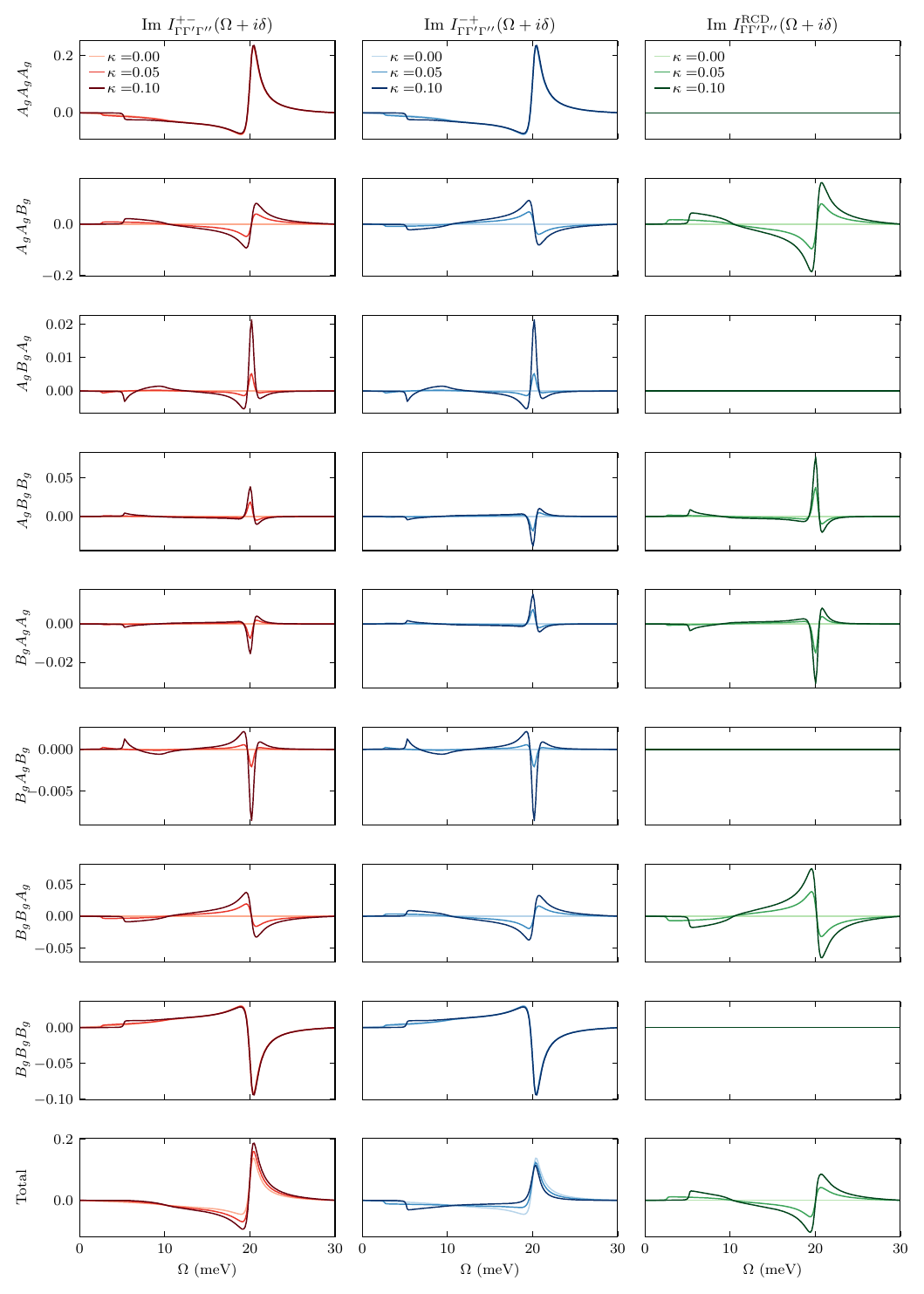}  %
\caption{\label{fig:right_pp_fano_contributions} \textbf{Irrep resolved Fano contributions} Right Fano-type diagram [Fig.~\ref{fig:diagram_panel}(h)] for the particle-particle channel $I^{ss^\prime}_{\text{Fano, pp,R}}(\Omega)$. Columns show different polarization channels: (1) $\bm{e}^+$ in, $\bm{e}^-$ out; (2) TRS-reversed process; (3) their difference (RCD). Rows correspond to different combinations of irreps for the photon-phonon, spin bilinear, and photon-spin vertices. We show the imaginary part of the analytical continued Matsubara Raman correlator, which is nonzero only if the initial and final irreps differ. Evaluated with the parameters from Tab.~\ref{tab:model_parameters}. } 
\end{figure*}

Finally, we consider the contribution to the RCD originating from the \textit{Fano-type} processes. The corresponding Raman intensities are shown diagrammatically in Fig.~\ref{fig:diagram_panel}~(g) and (h), which depict the diagrams with the polarization bubble attached to the left and right Raman vertices, respectively.
These contributions,
$I^{+-}_{\mathrm{Fano,L/R}}(\Omega)$ and $I^{-+}_{\mathrm{Fano,L/R}}(\Omega)$,
can be computed from the explicit expressions given in Eqs.~(\ref{IfanofinalL}) and (\ref{IfanofinalR}) of the {\it Methods} section.
Then we combine the two vertices by noting that, in the sum over irreducible representations, the vertex of the left diagram appears as the complex conjugate of that of the right diagram.
Using the same general Raman tensors as before, the resulting RCD contribution for $\Omega > 0$ takes the form
\begin{align} 
&I^{\text{RCD}}_{\text{Fano,R+L}}(\Omega) = 
    2 \, \text{Im} \int 
    \frac{\mathrm{d}^2 k}{(2\pi)^2} \sum_{\substack{\Gamma, \Gamma^\prime, \Gamma^{\prime \prime} }}
    \left[ \frac{D_{\Gamma^\prime \Gamma^{\prime \prime}} (\Omega) }{\Omega + i\delta - 2 E_{\bm k}} \right] \nonumber  \\ 
    &\times 2 \, \text{Re}\bigg\{ i \, \Lambda^*_{\Gamma^\prime}(\bm k ) 
    \Big[ \big( r^{bb}_{\text{em-sp},\Gamma}(\bm k ) - r^{aa}_{\text{em-sp},\Gamma}(\bm k ) \big) 
    \, r^{ab}_{\text{em-ph},\Gamma^{\prime \prime}} 
    - \big( r^{bb}_{\text{em-ph},\Gamma^{\prime \prime}} - r^{aa}_{\text{em-ph},\Gamma^{\prime \prime}} \big) 
    \, r^{ab}_{\text{em-sp},\Gamma}(\bm k )  \Big] \bigg\}.
\end{align}
This expression highlights the two key conditions, which must be satisfied in order to obtain a finite RCD. 
First, there must be an overlap between the phonon mode and the chiral spin continuum, enabling hybridization through spin–phonon interactions in analogy with the Fano effect. 
Second, the Raman response must either allow mixing between different irreps or involve both diagonal ($r^{\mu\mu}$) and off-diagonal ($r^{\mu\nu}, \, \mu\neq\nu$) components of the Raman tensor. 
In particular, the symmetry constraints of the $C_{2h}$ point group dictate that a finite RCD can only arise from mixing between different irreps.
In Fig.~\ref{fig:right_pp_fano_contributions} we demonstrate the mixing condition by showing the irrep-resolved contributions of the right Fano diagram Im $I_{\Gamma \Gamma^\prime \Gamma^{\prime\prime}}^{ss^\prime}(\Omega +i\delta)$ in the $+-$ and $-+$ scattering channels, together with the corresponding RCD. Each term $I_{\Gamma \Gamma^\prime \Gamma^{\prime\prime}}^{ss^\prime}(\Omega+i\delta)$  isolates the contribution in which the photon couples first to a phonon of symmetry  $\Gamma$, then mixes through a spinon intermediate state of symmetry $\Gamma'$, and finally returns through a phonon channel.
The contribution $I_{\Gamma \Gamma^\prime \Gamma^{\prime\prime}}^{\text{RCD}}(\Omega+i\delta)$ is only finite, if  the phonon–photon and spinon–photon vertices belong to different irreps, as expected from the $C_{2h}$ symmetry constraints. Therefore, a finite RCD originates from inter-irrep mixing, which couples light to the TRS-breaking hybridized phonons $A_g \pm i B_g$, giving rise to the observed chiral response.

\begin{table}[h]
    \centering
    \caption[Model Parameters]{\label{tab:model_parameters} Coupling constant, bare phonon frequency and phonon Raman matrix elements used to evaluate the Raman circular dichroism shown in Fig. \ref{fig:rcd_overview} and the irrep resolved right Fano contributions in the particle-particle channel Fig.~\ref{fig:right_pp_fano_contributions}. }
    \begin{tabular}{l l c}
        \toprule
        \textbf{Symbol} & \textbf{Description} & \textbf{Value} \\
        \midrule
        $\nu$ & Spin-photon coupling                 & $-0.12$ \\
        $r^{aa}$                             & Components of $r_{\text{em-ph}, A_g}$ = \text{diag}$(r^{aa}_\text{em-ph},-r^{aa}_\text{em-ph})$            & $-2.00$ \\
        $r^{ab}$                             & Components of $r_{\text{em-ph},B_g} =\begin{pmatrix} 0 & r^{ab}_\text{em-ph} \\ r^{ab}_\text{em-ph} & 0 \end{pmatrix}$        & $0.80$ \\
        $\omega_0$            & Bare phonon frequency     & $7.96 J$ \\
        $\lambda_{A_g} = \lambda_{B_g}$  &  Spin-phonon coupling   & $0.40$ \\
        $\mu_{A_g}$ = $\mu_{B_g}$ & Phonon-photon coupling & $0.03$ \\
    \bottomrule
    \end{tabular}
\end{table}

Fig.~\ref{fig:rcd_overview} illustrates the characteristic behavior of the RCD for a spin–phonon–coupled Kitaev spin liquid.
The calculations use the bare phonon frequencies and coupling parameters listed in Tab.~\ref{tab:model_parameters}\footnote{Note, that our model is written in terms of Pauli matrices, therefore the coupling constant $J_{\text{Pauli}}$ used is $1/4$ of the coupling $J_{\text{Spin 1/2}}$ of the Hamiltonian expressed in terms of spin operators, so $J_{\text{Pauli}}\approx 2.5$ meV.}.
 As we discussed above,
 we set the bare $A_g$ and $B_g$ phonons to have identical frequencies, $\omega^0_{A_g} = \omega^0_{B_g}=\omega_0$,
 and  spin-phonon couplings, so that the phonon splitting vanishes in the $\kappa \to 0$ limit.  The panels in Fig.~\ref{fig:rcd_overview} shows the individual Raman responses from spinons, phonons, and Fano interference, along with their total sum.
This decomposition reveals the distinct microscopic origins of the RCD and its dependence on the chiral parameter $\kappa$.
At $\kappa=0$, the circular channels remain degenerate and the RCD vanishes, while finite $\kappa$ generates a chiral spinon background that hybridizes with the phonons, leading to mode splitting and a finite, field-dependent RCD.

The phonon and Fano contributions are of comparable magnitude, and their relative weight can be tuned by the microscopic Raman tensor elements.
For example, when $r^{aa} = -r^{ab}$, the Fano contribution to the RCD cancels (even though the underlying Fano asymmetry in the total Raman intensity remains finite), leaving only the phonon contribution. Hence, the RCD provides a direct measure of intrinsic phonon chirality, distinguishing it from apparent asymmetries that can arise trivially from the analytic continuation of the Raman susceptibility.
In contrast, the spinon RCD and the field-dependent peak arising from Fano interference contribute with 
similar magnitude but opposite sign. 
In the total response, these two nearly cancel, such that the intrinsic spinon RCD and in particular its field dependence, is strongly masked in experimentally accessible spectra.

\subsection*{Comparison to experiment}
\label{sec:comparsion_to_exp}

Having established the microscopic origin of Raman circular dichroism in the spin–phonon coupled Kitaev model, we now compare our predictions with experimental data on $\alpha$-RuCl$_3$ reported by Sahasrabudhe \textit{et al.}~\cite{sahasrabudhe2024,Sahasrabudhethesis2024}. Fig.~\ref{fig:intro_figure}~(b) shows the result of this comparison, with the theoretical curves obtained using the parameters listed in Tab.~\ref{tab:thyvsexp_parameters}.
The helicity-resolved experimental spectra $I^{+-}$  (red dashed line with circles indicating the data points) and $I^{-+}$ (blue dashed line with circles indicating the data points) display a pronounced amplitude asymmetry together with a small yet discernible shift in their peak positions. These two features, the intensity imbalance and the frequency offset, are precisely the quantities that determine the magnitude and the sign of the observed RCD.
To capture both features, we allow the parameters in
Tab.~\ref{tab:thyvsexp_parameters} to differ slightly between channels. This accounts for the fact that the frequencies entering the Fano diagrams correspond to renormalized phonons, which are naturally split once the TRS-breaking self-energy is included.

The numerical evaluation of the Raman response is performed using the same $\mathbf{k}$-space sampling and broadening parameters as in Fig.~\ref{fig:rcd_overview}, ensuring direct consistency with our earlier calculations. For the comparison with experiment, we refine only the energy resolution: the frequency integrals are evaluated on a modified grid covering the relevant phonon window, $\Omega \in [7.2J, 8.82J]$, discretized into 200 intervals. This finer grid captures the narrow phonon features and the subtle peak shifts that are essential for resolving the helicity-dependent asymmetry in the experimental spectra.

Although our fitting procedure allows us to reproduce these qualitative features, we emphasize that the minimal Kitaev model cannot yield a fully quantitative description of $\alpha$-RuCl$_3$. A realistic treatment requires incorporating the substantial Heisenberg and $\Gamma$ interactions known to be present in the material \cite{Maksimov2020,Moller2025}. These additional exchanges reshape the magnetic continuum and thus influence the detailed Raman response, making exact agreement unattainable within the simplified model used here. Moreover, in our calculations the effect of the magnetic field is modeled through the effective three-spin term $\kappa$, which corresponds to the leading-order expansion of a field applied along the high-symmetry $[111]$ direction. In the experiment, however, the field is applied along the crystallographic $a$-axis, for which the projection onto the spin-orbit coupled pseudospin basis differs from the ideal $[111]$ case and gives a larger weight to the out-of-plane ($z$) component. Consequently, the relationship between the experimental field strength and the effective $\kappa$ used in the model is only approximate. Finally, the slight energy splitting between the $A_g$ and $B_g$ phonons causes them to couple to different regions of the magnetic continuum, which further enhances the asymmetry observed in the helicity-resolved Raman spectra.

Nevertheless, the essential features of the experimental response are captured: the low-lying $A_g$ phonon near 20 meV shows a pronounced dichroism between the $I^{+-}$ and $I^{-+}$ channels, and its Fano-like asymmetry is consistent with hybridization between the phonon and the magnetic continuum \cite{sahasrabudhe2024,Sahasrabudhethesis2024}. Our theoretical framework reproduces these characteristic signatures and clarifies the microscopic origin of the observed RCD.

\begin{table}[h]
    \centering
    \caption[Theory vs Experiment Parameters]{\label{tab:thyvsexp_parameters} 
    Coupling constant, bare phonon frequencies and phonon Raman matrix elements used to evaluate the Raman circular dichroism shown in Fig.~\ref{fig:intro_figure} comparing theoretical and experimental data.}
    \begin{tabular}{l l c}
        \toprule
        \textbf{Symbol} & \textbf{Description} & \textbf{Value} \\
        \midrule
        $\nu$ & Spin-photon coupling                 & $-0.08$ \\
        $r_{aa}$                             & Components of $r_{\text{em-ph},A_g}$ = \text{diag}$(r^{aa}_\text{em-ph},-r^{aa}_\text{em-ph})$            & $-1.40$ \\
        $r_{ab}$                             & Components of $r_{ \text{em-ph},B_g} =\begin{pmatrix} 0 & r^{ab}_\text{em-ph} \\ r^{ab}_\text{em-ph} & 0 \end{pmatrix}$        & $1.20$ \\
        $\omega_{A_g}^0$            & Bare phonon frequency, $A_g$ mode      & $7.95J$ \\
        $\lambda_{A_g}$  & $A_g$ Spin-phonon coupling   & $0.25$ \\
        $\mu_{A_g}$ & $A_g$ Phonon-photon coupling  & $0.031$ \\
       $\omega_{B_g}^0$            & Bare phonon frequency, $B_g$ mode    & $8.08J$ \\
         $\lambda_{B_g}$  & $B_g$ Spin-phonon coupling   & $0.38$ \\
        $\mu_{B_g}$ & $B_g$ Phonon-photon coupling & $0.042$ \\
        \bottomrule
    \end{tabular}
\end{table}

\section*{Discussion}
In this paper, we discussed the microscopic origin and manifestations of chiral phonons in a spin–phonon coupled Kitaev system. While in non-centrosymmetric crystals such modes can arise already at the single-particle level, where the absence of inversion or mirror symmetries permits complex phonon eigenvectors carrying quantized angular momentum $\pm\hbar$, here we showed that even in centrosymmetric systems phonon chirality can emerge dynamically from many-body interactions. In particular, coupling between phonons and fractionalized spin excitations generates complex off-diagonal components of the phonon self-energy, which hybridize orthogonal linear modes into eigenstates of opposite helicity. This interaction-driven mechanism provides a natural route for the emergence of chiral phonons in otherwise achiral lattices.

In particular, we showed that hybridization between phonons and the Majorana continuum in the TRS broken phase of the Kitaev model induces phonon chirality, which results in a characteristic RCD signature and leads to several key findings. The RCD originates from the interference between spinon, phonon, and Fano channels in the Raman response, where the complex self-energy endows the mixed phonon modes with opposite helicities that couple differently to left- and right-circularly polarized light. The resulting imbalance in Raman intensities provides a direct spectroscopic fingerprint of interaction-induced chiral phonons and establishes Raman spectroscopy as a sensitive probe of their evolution across TRS breaking regimes.

Finally, we apply our theoretical framework to the prototypical Kitaev material $\alpha$-RuCl$_3$, where recent polarized Raman measurements~\cite{Sahasrabudhethesis2024} have revealed signatures consistent with interaction-induced phonon chirality. Comparison with experiment shows qualitative agreement: the calculated RCD reproduces both the observed sign change near 20 meV and its magnetic-field dependence, vanishing in the $\kappa=0$ limit and increasing monotonically with field strength. At higher energies (20.2–20.8 meV), the minimal Kitaev model slightly over- and underestimates the shoulder features of the $I^{+-}$ and $I^{-+}$ channels. This discrepancy stems from the oversimplification  of the pure Kitaev honeycomb model with the three-spin  TRS breaking term, as opposed to more realistic descriptions \cite{Winter2017,Maksimov2020,Moller2025}.

More broadly, the mechanism of interaction-induced phonon chirality introduced here is generic and applies to a wide range of systems with spin–phonon coupling, including those hosting magnons or spinons. In essence, the effect originates from the lowering of crystal symmetry when TRS is broken. This symmetry reduction allows previously independent phonon modes to mix through complex off-diagonal terms in their self-energy, giving rise to chiral lattice vibrations. 
Beyond the Majorana–phonon hybridization analyzed in this work, magnon–phonon coupling has been experimentally observed in the layered zigzag antiferromagnet FePSe$_3$, in quantitative agreement with first-principles calculations~\cite{Cui2023}. Moreover, spin–orbit coupling alone can mediate phonon–magnon hybridization, giving rise to chiral phonons even in the absence of explicit magnetic order~\cite{Ren2024}.

\section*{Methods}

In the following we set up the diagrammatic framework used to compute the RCD of a spin–phonon coupled Kitaev model. We first evaluate the one-loop polarization bubble which captures the effect of the interaction under the RPA approximation.  
We then calculate the Matsubara Raman correlator~$\mathcal{I}(i\Omega,\beta)$ and specify the photon–spin and photon–phonon Raman vertices and evaluate the response in the interaction picture. Finally, we separate the total Raman signal into three contributions: pure spinon scattering, renormalized phonon scattering, and Fano interference. The Raman intensity Eq.~\eqref{eqn:intensityram} follows from analytic continuation of the Matsubara correlator Eq.~\eqref{eq:matsubaraRaman} and the fluctuation-dissipation theorem 
\begin{align}
\label{eq:FDT}
    I(\Omega,\beta ) = \frac{2}{1-e^{-\beta \Omega}} \text{Im} \mathcal{I}(i\Omega,\beta)\big|_{i\Omega \to \omega + i\delta}, 
\end{align}
where in the Matsubara Raman correlator we sum over the internal irrep indices.

\subsection*{Polarization bubble}
\label{app:polarization_bubble}

In the $C_{2h}$ point group, 
the spin bilinears entering the spin–phonon Hamiltonian, Eq.~\eqref{eq:spin-phonon-coupling}, describe the coupling to the $A_g$ and $B_g$ phonons originating from the doubly degenerate $E_g$ modes of the parent $D_{3d}$ group.
Their explicit forms are given in Eqs.~\eqref{sigmaA} and \eqref{sigmaB}.
Expressed in the quasiparticle  basis used to diagonalize the extended Kitaev model [Eq.~(\ref{spindiag})], these bilinears take the matrix form
 \begin{align}
\label{app:khm_raman_operator_in_complex_fermions}
    \Sigma_{\Gamma}(\bm k)  = 
    \begin{pmatrix} a^\dagger_{\bm k}  a_{-\bm k }  \end{pmatrix}
    \begin{pmatrix}
         \Lambda_{\Gamma,11}(\bm k) &  \Lambda_{\Gamma,12}(\bm k) \\
          \Lambda_{\Gamma,21}(\bm k) &  \Lambda_{\Gamma,22} (\bm k)
    \end{pmatrix} 
    \begin{pmatrix} a_{\bm k} \\ a^\dagger _{-\bm k }  \end{pmatrix}
\end{align}
with $ \Lambda_{\Gamma,22}(\bm k) =  - \Lambda_{\Gamma,11} (\bm k)$ and $ \Lambda_{\Gamma,12} (\bm k)= \Lambda_{\Gamma,21}^* (\bm k)$.  The matrix elements are given by
\begin{align}
    \Lambda_{\Gamma,11}(\bm k) =& \text{Re}\, F_{\Gamma}(\bm k) \cos2\theta_{\bm k } - \text{Im}\, F_{\Gamma}(\bm k)  \sin2\theta_{\bm k } \sin\phi_{\bm k }\\
    \Lambda_{\Gamma, 12}(\bm k) =& i\,
    \text{Im}\, F_{\Gamma}(\bm k)  \left( \cos^2\theta_{\bm k } + e^{2i\phi_{\bm k }}\sin^2\theta_{\bm k }\right)+ 
    e^{i\phi_{\bm k }}\text{Re}\, F_{\Gamma}(\bm k)  \sin(2\theta_{\bm k }) 
\end{align}
 where the Bogoliubov angle $\theta_{\bm k}$ is defined as $\tan 2 \theta_{\bm k }=\frac{|\Delta_{\bm k }| }{\xi_{\bm k }}$ with
    $\xi_{\bm k } = \text{Re}  S(\bm k )$, $\Delta_{\bm k } = - \kappa(\bm k ) - i \text{Im} S(\bm k )$ with $
    S(\bm k ) = \sum_{\alpha} J_{\alpha} e^{i \bm k \cdot \bm d_\alpha}$ and $
     \kappa(\bm k ) = - 2 K \sum_{\alpha'} \sin( \bm k \cdot \bm  d_{\alpha'})$ \cite{Kitaev2006}. Here, the sums over $\alpha$ and $\alpha'$ run over the nearest- and second-neighbors, respectively.
The form factors $F_{\Gamma}(\bm k)$ depend on the irreps as
\begin{align} 
F_{A_{g}} (\bm k) &=  e^{i \bm k \cdot \bm d_x} + e^{i \bm k\cdot \bm d_y} - 2 e^{i \bm k \cdot \bm d_z},  \\ 
F_{B_g} (\bm k) &=  \sqrt{3} \left( e^{i \bm k \cdot \bm d_x} - e^{i \bm k\cdot \bm d_y} \right)
\end{align} 
At zero temperature only $\Lambda_{\Gamma,12}(\bm k )$ contributes to Stokes scattering and we use the abbreviation  $\Lambda_{\Gamma}(\bm k ) =  \Lambda_{\Gamma,12} (\bm k )=  \Lambda_{\Gamma,21}^*(\bm k )$.

The spin bilinears $\Sigma_{\Gamma}(\bm k)$ define the structure of the spin–phonon interaction in the  momentum space and determine how lattice vibrations couple to the Majorana quasiparticles of the  Kitaev spin liquid~\cite{Ye_2020}.
These bilinears enter directly into the one-loop correction to the phonon propagator through the polarization bubble, which captures the virtual excitation of pairs of Majorana fermions that renormalize the phonon dynamics.
We therefore compute the corresponding one-loop correction to the phonon propagator arising from the spin–phonon coupling, Eq.~\eqref{eq:spin-phonon-coupling}.
The Dyson equation takes the form

\begin{equation}
\begin{adjustbox}{valign=c}
\begin{tikzpicture}[scale=1.2]
    \draw[white] (0,0) -- (0.7,0.7);
    \draw[white] (0,0) -- (0.7,-0.7);
    
    \draw[double, dash pattern=on 3pt off 3pt] ( -1.6,0) -- (-0.8,0);
    \node at (-0.6,0) { $=$ };
    \draw[-, dash pattern=on 3pt off 3pt] (-0.4,0) to (0.4,0);
    \node at (0.6 ,0) { $+$ };
    \draw[-, dash pattern=on 3pt off 3pt] (0.6,0) to (1.4,0);
    \draw[-] (1.4,0) to[bend left=55] (2.6,0);
    \draw[-] (2.6,0) to[bend left=60] (1.4,0);
    \draw[double, dash pattern=on 3pt off 3pt] (2.6,0) -- (3.4,0);
    \filldraw[black] (1.4, 0) circle (2pt);
    \filldraw[black] (2.6, 0) circle (2pt);
\end{tikzpicture}
\end{adjustbox}
\label{fig:Dyson}
\end{equation}

with the polarization bubble given in imaginary time as $\Pi(\tau) = -\langle T_\tau H_\text{sp-ph}(\tau) H_\text{sp-ph}(0) \rangle$. 

\begin{figure}[t]
    \centering
    \includegraphics[width=1\linewidth]{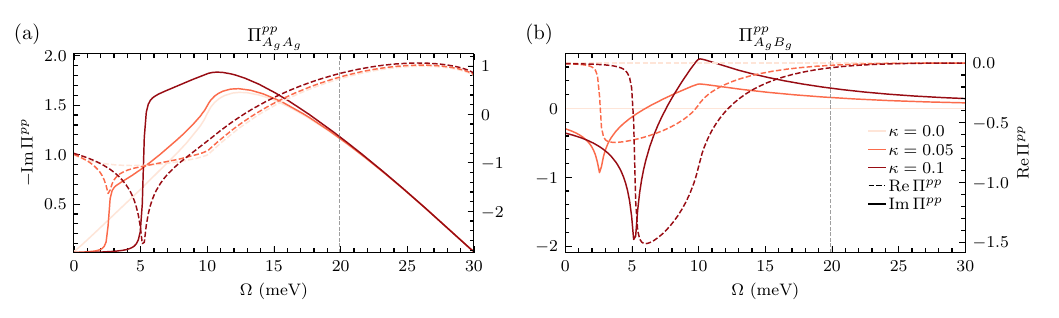}
    \caption{\textbf{Polarization bubble $\Pi_{\Gamma,\Gamma^\prime}(\Omega,0)$} Imaginary part red gradient solid, real part red gradient dashed. (a) Diagonal component of the polarization bubble (b) off-diagonal component of the polarization bubble. Note $\Pi_{A_g A_g}(\Omega,q) = \Pi_{B_g B_g}(\Omega,q)$ and $\Pi_{A_g B_g}(\Omega,\bm q) = - \Pi_{B_g A_g}(\Omega,\bm q)$. At $\kappa=0$ the off-diagonal polarization bubble vanishes. The gray line indicates the energy of the bare $E_g$ doublet.}
    \label{fig:polarization_bubble}
\end{figure}

Expressing $H_\text{sp-ph}$ in terms of Bogoliubov fermions, $\Pi$ contains particle-particle (pp), particle-hole (ph) and hole-hole (hh) contributions. However, due to the presence of only one physical band $E(\bvec k)$ and $\bvec{q}=0$ momentum transfer, the ph and channels vanishes
\begin{align}
\Pi^{pp}_{\Gamma \Gamma^\prime} (i\omega_m, \bm q)  &= - \lambda_\Gamma \lambda_{\Gamma^\prime}  \int \frac{\mathrm{d}^2 k }{(2\pi)^2} \Lambda_{\Gamma}(\bm k ) \Lambda_{\Gamma^\prime}^*(\bm k  ) \frac{1}{\beta}\sum_{i\nu_n}G(i\omega_m + i\nu_n,\bm q+ \bm k )G(-i\nu_n,-\bm k ) \\
&= \lambda_\Gamma \lambda_{\Gamma^\prime}  \int \frac{\mathrm{d}^2 k }{(2\pi)^2}
 \Lambda_{\Gamma}(\bm k  ) \Lambda_{\Gamma^\prime}^*(\bm k  ) \frac{f(E_{\bm q+ \bm k }) - f(-E_{-\bm k })}{E_{\bm q+ \bm k } + E_{-\bm k } - i \omega_m} \\
\Pi^{hh}_{\Gamma \Gamma^\prime} (i\omega_m,\bm q) & =-\lambda_\Gamma \lambda_{\Gamma^\prime}  \int \frac{\mathrm{d}^2 k }{(2\pi)^2} 
  \Lambda_{\Gamma}^*(\bm k ) \Lambda_{\Gamma^\prime}(\bm k )  \frac{1}{\beta}\sum_{i\nu_n}G(-i\omega_m - i\nu_n,\bm q+ \bm k )G(i\nu_n,- \bm k ) \\
&=  \lambda_\Gamma \lambda_{\Gamma^\prime}  \int \frac{\mathrm{d}^2 k }{(2\pi)^2} \Lambda_{\Gamma}^*(\bm k ) \Lambda_{\Gamma^\prime}(\bm k ) \frac{f(E_{\bm q+ \bm k }) - f(-E_{-\bm k })}{E_{\bm q+\bm k } + E_{-\bm k } + i \omega_m} ,
\end{align}
where the total polarization bubble is the sum $\Pi_{\Gamma \Gamma^\prime} (i\omega_m) = \Pi^{pp}_{\Gamma \Gamma^\prime} (i\omega_m) + \Pi^{hh}_{\Gamma \Gamma^\prime} (i\omega_m)$. The analytic continuation $i\omega_m \rightarrow \omega + i\delta$ gives access to the retarded real-frequency polarization bubble. 
Fig.~\ref{fig:polarization_bubble} shows $\text{Re}\Pi^{pp}_{\Gamma \Gamma^\prime} $ and $\text{Im}\Pi^{pp}_{\Gamma \Gamma^\prime}$.

\subsection*{Raman intensity in spin-phonon coupled systems}
\label{app:Theory_Raman_spc_systems}

We now derive the Raman intensity by evaluating the relevant diagrams in the Matsubara formalism and performing a systematic expansion in the spin-phonon interaction. The spin–phonon coupled Raman intensity $I(\Omega,\beta)$ Eq.~\eqref{eqn:intensityram} relates via Eq.~\eqref{eq:FDT} to the Matsubara Raman correlator
\begin{align}
\label{eq:matsubaraRaman}
     \mathcal{I}^{ss^{\prime} }(i \Omega,\beta)  &= - \int_{0}^\beta \mathrm{d}\tau  e^{i \Omega \tau} \Big \langle \mathcal{T}_\tau 
         \left[R^{s s^\prime}(\tau)\right]^\dagger R^{s s^\prime}(0) e^{-i \int_0^\beta \mathrm{d}\tau^\prime H_\text{sp-ph}(\tau^\prime)} \Big\rangle,
\end{align}
with the $S$-matrix  defined as $S(\beta) = \mathcal{T}_\tau \exp\left( - \int_0^\beta \mathrm{d}\tau H_\text{sp-ph} (\tau) \right)$.  The $S$-matrix can be systematically expanded in powers of the interaction Hamiltonian. This perturbative expansion produces a series of terms that represent successive orders of spin–phonon processes. Explicitly, one finds
 \begin{eqnarray}
       \mathcal{I} ^{s s^{\prime}  }(i \Omega,\beta)  
        &&= - \int_{0}^\beta e^{i \Omega \tau} \left \langle \mathcal{T}_\tau 
         \left[ R^{s s^\prime}(\tau)\right]^\dagger  R^{s s^\prime }(0) 
          \exp\left( - \int_0^\beta \mathrm{d}\tau H_\text{int} (\tau) \right)\right \rangle /\left\langle S \right\rangle \nonumber\\
         &&= - \sum_{k=0}^{\infty}(-1)^k \int_{0}^\beta \prod_{i=0}^k \mathrm{d}\tau 
         \left \langle 
         \mathcal{T}_\tau 
         \left[ R^{s s^\prime}(\tau)\right]^\dagger  R^{s s^\prime }(0)\prod_{i=0}^k H_{sp-ph}(\tau_i) 
         \right \rangle.
\end{eqnarray} 
Here the denominator $\langle S \rangle$ cancels disconnected contributions according to the linked-cluster theorem, ensuring that only connected diagrams contribute to the Raman response.
We expand the series using Eqs.~\eqref{eq:def_phonon_R} and \eqref{eq:def_LF_R}, retaining terms up to one loop in RPA.
 The resulting contributions are represented diagrammatically in Fig.~\ref{fig:diagram_panel}~(d)–(h). Each diagram corresponds to a distinct physical processes and we evaluate them below.
 
The \textit{spinon Raman contribution} is represented diagrammatically 
in Fig.~\ref{fig:diagram_panel} (d)  and  is given by
\begin{align}
    \mathcal{I}_\text{sp}^{ss^\prime} (i\Omega,\beta) = - \int \frac{\mathrm{d} \bm k}{(2\pi)^2} &
    \nu^2
     \bigg[ \sum_\Gamma \sum_{\mu ,\nu}  r_{\text{em-sp},\Gamma}^{\mu \nu}(\bm k)
     e^{*s}_{\mu}  e^{s^\prime}_{\nu} \bigg] 
     \bigg[
     \sum_{\Gamma^\prime} 
      \sum_{\mu^\prime ,\nu^\prime} 
      r_{\text{em-sp},\Gamma^{\prime}}^{*\mu^\prime \nu^\prime}(\bm k) 
     e^{s}_{\mu^\prime}  e^{*s^\prime}_{\nu^\prime} \bigg]\nonumber\\
     \times \frac{1}{\beta} &\sum_{\omega_n} G^0(i\omega_n + i\Omega,\bm k) G^0(-i \omega_n,-\bm k),
\end{align}
 where  $G^0(i\omega_n + i\Omega,\bm k)$ denotes the bare  Majorana fermion Green's function Eq.~\eqref{eqn:greenfunc}. After  Matsubara summation, analytical continuation and with the FDT we obtain the Raman response from the spinons 
\begin{align}\label{Ispinfinal}
    I_\text{sp}^{ss^\prime}(\Omega,\beta) = \int \frac{\mathrm{d} \bm k}{(2\pi)^2} &
    \nu^2
     \left[ \sum_\Gamma \sum_{\mu ,\nu}  r_{\text{em-sp},\Gamma}^{\mu \nu}(\bm k)
     e^{*s}_{\mu}  e^{s^\prime}_{\nu} \right] 
     \left[
     \sum_{\Gamma^\prime} 
      \sum_{\mu^\prime ,\nu^\prime} 
      r_{\text{em-sp},\Gamma^{\prime}}^{*\mu^\prime \nu^\prime}(\bm k) 
     e^{s}_{\mu^\prime}  e^{*s^\prime}_{\nu^\prime} \right]\nonumber\\
     &\times 2\pi \left[ \delta\left(\Omega - 2E_{\bm k }\right)  \big[ 1 - n_f(E_{\bm k} )\big]^2 +\delta\left(\Omega + 2 E_{\bm k} \right)\big[ n_f(E_{\bm k} )\big]^2 \right]. 
\end{align}
The Anti-Stokes peak proportional to $\big[ n_f(E_{\bm k} )\big]^2$ from the hh-chanel does not contribute at $T=0$. 

The  \textit{phonon Raman intensity}, represented by the diagram in Fig.~\ref{fig:diagram_panel}(e), is given by
\begin{align}
\mathcal{I}^{ss^\prime}_\text{ph}(i\Omega,\beta) = -\int_{0}^\beta \mathrm{d}\tau e^{i \Omega \tau} 
         \left\langle\mathcal{T}_\tau 
          \left[\sum_{\Gamma} \mu_\Gamma \sum_{\mu ,\nu} e^{s}_\mu  e^{*s^\prime}_\nu r_{\text{em-ph},\Gamma}^{\mu \nu} u_{\Gamma}(\tau)\right]
          \left[\sum_{\Gamma^\prime} \mu_{\Gamma^\prime} \sum_{\mu ,\nu} e^{s}_\mu  e^{*s^\prime}_\nu r_{\text{em-ph}, \Gamma^\prime}^{\mu \nu} u_{\Gamma^\prime}(0)\right]
        \right\rangle .
\end{align}
In the bare phonon case [Eq.~\eqref{D0}],  when $D^0_{\Gamma \Gamma^\prime} \sim D^0_{\Gamma}\delta_{\Gamma \Gamma\prime}$, we get
\begin{align} 
\mathcal{I}_\text{ph}^{ss^\prime}(i\Omega,\beta) = 
\sum_{\mu,\nu,\mu^\prime\nu^\prime} \sum_{\Gamma}
\left[ r^{\mu\nu}_{\text{em-ph},\Gamma} e^{s}_\mu  e^{*s^\prime}_\nu  \right]^\dagger  r^{\mu^\prime \nu^\prime }_{\text{em-ph},\Gamma} e^{s}_{\mu^\prime}  e^{*s^\prime}_{\nu^\prime}  D^0_\Gamma(i\Omega).
\end{align}
For $T\to0$, the Raman intensity consists of sharp peaks at the bare phonon energies
\begin{align}
        I_\text{ph}^{ss^\prime}(\Omega) = 2 \sum_{\mu,\nu,\mu^\prime\nu^\prime} \sum_{\Gamma} \mu_{\Gamma}^2 
        \left[ r^{\mu\nu}_{\text{em-ph},\Gamma} e^{s}_\mu  e^{*s^\prime}_\nu  \right]^\dagger  r^{\mu\nu}_{\text{em-ph},\Gamma} e^{s}_\mu  e^{*s^\prime}_\nu
        \Big[
        \left[ 1+n_B(\omega_\Gamma) \right] \delta(\Omega - \omega_\Gamma) +
        n_B(\omega_\Gamma) \delta(\Omega + \omega_\Gamma)
        \Big].
\end{align}
However, if we consider all the bubble diagrams in RPA resummation, the resulting full phonon propagator allows for mixing between different irreps. The intensity in this renormalized case is
\begin{align}\label{Iphononfinal}
     I^{ss\prime}_{\substack{\text{ph}}} (\Omega,\beta) = \frac{-2}{1-e^{-\beta\Omega} }  \text{Im} \sum_{\Gamma, \Gamma^\prime } \mu_{\Gamma}\mu_{\Gamma^\prime} 
     \sum_{\mu,\nu,\mu^\prime\nu^\prime} 
        \left[ r^{\mu\nu}_{\text{em-ph},\Gamma} e^{s}_\mu  e^{*s^\prime}_\nu  \right]^\dagger  r^{\mu^\prime \nu^\prime }_{\text{em-ph},\Gamma^\prime} e^{s}_{\mu^\prime} e^{*s^\prime}_{\nu^\prime}  D_{\Gamma \Gamma^\prime} (\Omega+i\delta).
\end{align}

Finally, the  \textit{Fano Raman  contributions}, originating from diagrams Fig.~\ref{fig:diagram_panel}~(g) and (h), are 
{\small\begin{align} 
\mathcal{I}^{ss\prime}_{\substack{\text{Fano,L}}}(i\Omega,\beta)=\int_0^\beta \mathrm{d}\tau\mathrm{d}\tau_1 e^{i \Omega \tau} 
         \left\langle\mathcal{T}_\tau 
          \left[\sum_{\Gamma} \alpha_\Gamma r^{s s^\prime }_{\text{em-sp},\Gamma} \Sigma_\Gamma(\tau)\right]
         \left[\sum_{\Gamma^\prime} \lambda_{\Gamma^\prime} \Sigma_{\Gamma^\prime}(\tau_1)u_{\Gamma^\prime}(\tau_1)\right]
         \left[\sum_{\Gamma^{\prime\prime}} \mu_{\Gamma^{\prime\prime}} \sum_{\mu ,\nu} e^{s}_\mu  e^{*s^\prime}_\nu r_{\text{em-ph},\Gamma^{\prime\prime}}^{\mu \nu} u_{\Gamma^{\prime\prime}}(0) \right]
         \right\rangle,
\end{align} }
{\small
\begin{align} 
         \mathcal{I}^{ss\prime}_{\substack{\text{Fano,R}}}(i\Omega,\beta)= \int_0^\beta \mathrm{d}\tau\mathrm{d}\tau_1 e^{i \Omega \tau} 
         \left\langle \mathcal{T}_\tau 
        \left[  \sum_{\Gamma} \mu_\Gamma \sum_{\mu ,\nu} e^{s}_\mu  e^{*s^\prime}_\nu r_{\text{em-p},\Gamma}^{\mu \nu} u_{\Gamma}(\tau) \right]
        \left[ \sum_{\Gamma^\prime} \lambda_{\Gamma^\prime} \Sigma_{\Gamma^\prime}(\tau_1)u_{\Gamma^\prime}(\tau_1) \right]
        \left[ \sum_{\Gamma^{\prime\prime}} \alpha_{\Gamma^{\prime\prime}} r^{s s^\prime }_{\text{em-sp},\Gamma^{\prime\prime}} \Sigma_{\Gamma^{\prime\prime}}(0) \right]
        \right\rangle.
\end{align} }
Performing the Matsubara summation we get the Fano contributions for $\Omega \geq 0$ at $T=0$:
\begin{align}\label{IfanofinalL}
     &I^{ss^\prime}_{\substack{\text{Fano} \\ \text{pp,L}}}(\Omega) = 2 \text{Im} \int \frac{\mathrm{d} \bm k}{(2\pi)^2} \sum_{\substack{\Gamma, \Gamma^\prime, \Gamma^{\prime \prime}  \\\mu, \nu, \mu^\prime ,\nu^\prime }} \nu \lambda_{\Gamma^\prime} \mu_{\Gamma^{\prime \prime}}
     r^{\mu \nu }_{\text{em-sp},\Gamma}(\bm k) e^{s}_\mu  e^{*s^\prime}_\nu 
     \Lambda^*_{\Gamma^\prime}(\bm k)
       r_{\text{em-ph},\Gamma^{\prime\prime}}^{\mu^\prime \nu^\prime}
     e^{*s}_{\mu^\prime}  e^{s^\prime}_{\nu^\prime} \frac{-D_{\Gamma^\prime \Gamma^{\prime \prime}}(\Omega+i\delta )}{\Omega + i\delta - 2 E_{\bm k}}, \\ \label{IfanofinalR}  &I^{ss^\prime}_{\substack{\text{Fano} \\ \text{pp,R}}}(\Omega) = 2 \text{Im}
     \int \frac{\mathrm{d} \bm k}{(2\pi)^2} \sum_{\substack{\Gamma, \Gamma^\prime, \Gamma^{\prime \prime}  \\\mu, \nu, \mu^\prime ,\nu^\prime }} \nu \lambda_{\Gamma^\prime} \mu_{\Gamma}
      r_{\text{em-ph},\Gamma}^{\mu \nu}
     e^{s}_{\mu}  e^{*s^\prime}_{\nu}
     \Lambda_{\Gamma^\prime}(\bm k)
      r^{*\mu^\prime \nu^\prime }_{\text{em-sp},\Gamma^{\prime\prime}}(\bm k) e^{*s}_{\mu^\prime}  e^{s^\prime}_{\nu^\prime} 
     \frac{-D_{\Gamma^\prime \Gamma^{\prime \prime}}(\Omega + i \delta)}{\Omega+ i \delta - 2 E_{\bm k}}. 
\end{align}

\section*{Data availability}
The theoretical datasets generated during and/or analyzed during the current study are available from the corresponding author upon reasonable request via Zenodo \cite{zenodo}. 

\section*{Code availability}
The codes used during the current study are available from the corresponding author upon reasonable request via Zenodo \cite{zenodo}. 

\section*{Acknowledgments}

The authors thank  Anuja Sahasrabudhe and  Paul H. M. van Loosdrecht 
for initiating the experimental Raman spectroscopy study of  phonon modes in the field-induced phase of  $\alpha$-RuCl$_3$   and granting access to their experimental data before publication. We also thank  Girsh Blumberg,  Kexin Feng, Jonas Habel, Valentin Leeb and Peng Rao for valuable discussions.
E.K. and N.B.P. acknowledge support from the Institute for Advanced Study (IAS) of the Technical University of Munich. 
S.S. and N.B.P. were supported by the U.S. Department   of Energy, Office
 of Science, Basic Energy Sciences under Award No. DE-SC0018056. 
J.K.  acknowledges support from the Deutsche Forschungsgemeinschaft (DFG, German Research Foundation) under Germany’s Excellence Strategy–EXC–2111–390814868, TRR 360 – 492547816 and DFG grants No. KN1254/1-2 and No. KN1254/2-1, as well as the Munich Quantum Valley, which is supported by the Bavarian state government with funds from the Hightech Agenda Bayern Plus. J.K.  acknowledges support from the TUM-Imperial flagship partnership. 
J.K.  and N.B.P. thank the hospitality of Aspen Center for Physics, which is supported by National Science Foundation grant PHY-2210452.

\section*{Author Contributions}
J.K. and N.B.P. devised the project.
E.K. performed the many-body calculations and the numerical calculations. 
E.K. and S.S. performed the symmetry analysis. 
E.K, S.S, J.K. and N.B.P. contributed to the interpretation of the results and the writing of the paper.

\section*{Competing interests}
The authors declare no competing interests.

\bibliography{main}
\bibliographystyle{custom_naturemag}
\end{document}